\documentclass{IEEEtran}
\usepackage{cite}
\usepackage{amsmath,amssymb,amsfonts}
\usepackage{algorithmic}
\usepackage{graphicx}
\usepackage{textcomp}
\usepackage[format=hang,font={small,bf,color={black}}, textfont={color=black}]{caption}
\usepackage{tikz}
\usepackage{braket}
\usetikzlibrary{backgrounds,shadows.blur,fit,decorations.pathreplacing,shapes}
\def\BibTeX{{\rm B\kern-.05em{\sc i\kern-.025em b}\kern-.08em
    T\kern-.1667em\lower.7ex\hbox{E}\kern-.125emX}}
\usepackage{braket}    
\usepackage{qcircuit}

\usepackage{mathtools}
\usetikzlibrary{arrows.meta}

\usepackage{graphicx}
\usepackage{epstopdf}
\graphicspath{{imgs/} }
\usepackage{array}

\usepackage{bm}
\usepackage{amsmath,amssymb}
\allowdisplaybreaks[4] 
\usepackage{booktabs, makecell, longtable} 
\usepackage{subcaption}
\usepackage{diagbox}
\usepackage[linesnumbered,ruled,vlined]{algorithm2e}

\usepackage{array}
\usepackage{booktabs}
\usepackage{multirow} 
\newcommand{\tabincell}[2]{\begin{tabular}{@{}#1@{}}#2\end{tabular}}  

\usepackage{color}
    
\begin{document}

\definecolor{dblue}{rgb}{0.00, 0.00, 0.75}
\newcommand{\ahsan}[1]{[{\color{blue}Ahsan: #1}]}
\newcommand{\claire}[2]{[{\color{red}Claire: #1}]}

\title{Support Vector Machines on Noisy Intermediate Scale Quantum Computers}
\author{
  \IEEEauthorblockN{Jiaying Yang, Ahsan Javed Awan and Gemma Vall-Llosera}\\
  \IEEEauthorblockA{Department of Cloud Systems and Platforms, Ericsson Research, Sweden\\
   \{claire.j.yang, ahsan.javed.awan, gemma.vall-llosera\}@ericsson.com}
}

\maketitle

\begin{abstract}
Support vector machine algorithms are considered essential for the implementation of automation in a radio access network. Specifically, they are critical in the prediction of the quality of user experience for video streaming based on device and network-level metrics. Quantum SVM is the quantum analogue of the classical SVM algorithm, which utilizes the properties of quantum computers to speed up the algorithm exponentially. In this work, we derive an optimized preprocessing unit for a quantum SVM that allows classifying any two-dimensional datasets that are linearly separable. We further provide a result readout method of the kernel matrix generation circuit to avoid quantum tomography that, in turn, reduces the quantum circuit depth. We also derive a quantum SVM system based on an optimized HHL quantum circuit with reduced circuit depth.
\end{abstract}

\begin{IEEEkeywords}
quantum support vector machine, noisy intermediate scale quantum computers, HHL, algorithm
\end{IEEEkeywords}

%
\IEEEpeerreviewmaketitle

\section{Introduction}
\label{sec:introduction}
Machine learning (ML) algorithms such as support vector machine (SVM), K-Means or simple linear regression solvers are crucial functional blocks of the user-data and management planes of a radio access network. Support vector machine algorithms, for example,  are considered essential for the implementation of automation in a radio access network. Specifically, they are critical in the prediction of the quality of user experience for video streaming based on device and network-level metrics\cite{ahmed2018automated}. In a live network and because of the large amount of training data, the training process of supervised ML algorithms is usually very time-consuming, and these algorithms need to be executed in specific hardware. An example of hardware specificity is a processor that exploits the properties of quantum mechanics, i.e., a quantum computer. Currently available quantum machines, also named noisy intermediate scale quantum (NISQ) computers, bear the promise of better performance on many tasks than current classical computers, but the quantum noise and decoherence time of the qubits limits the size of quantum circuits that can be reliably executed \cite{preskill2018quantum}. Thus, there is a need to reduce the complexity of quantum algorithms to observe expected results on NISQ computers. This paper introduces a quantum SVM system that is suitable to be implemented on NISQ computers.

SVM is a supervised machine learning technique for solving classification problems. It can classify vectors into two-subgroups, but in order to execute the SVM algorithm on a quantum machine, the algorithm needs to be redesigned so that the properties of quantum mechanics are fully exploited. The quantum support vector machine (QSVM) can quadratically or exponentially speed up the original classical algorithm depending on the circuit design chosen. In this paper, we explore the QSVM algorithm, and we find that:

\begin{enumerate}
    \item Optimizing the preprocessing unit for the QSVM system allows the QSVM model to classify any two-dimensional datasets that are linearly separable by a line crossing the origin
    \item Using a classical result readout method of the training-data oracle allows us to avoid using the quantum tomography technique, and using a new training-data oracle allows us to reduce the quantum circuit depth for small-scale training data
    \item Proposing a new  Harrow/Hassidim/Lloyd (HHL) quantum circuit allows us to use a shorter-depth circuit and observe better results on real quantum computers, and redesigning the result readout method enables us to solve QSVM problems
\end{enumerate}

This paper is organized as follows: Section \uppercase\expandafter{\romannumeral2} reviews the SVM algorithm, NISQ computers and basic concepts of quantum computing;  Section \uppercase\expandafter{\romannumeral3} introduces the related works in this area; Section \uppercase\expandafter{\romannumeral4} provides the implementation method of our QSVM system; Section \uppercase\expandafter{\romannumeral5} introduces the datasets, platforms, metric and the baseline of our implementation; Section \uppercase\expandafter{\romannumeral6} shows the results of both our implementation and the prior art; and Section \uppercase\expandafter{\romannumeral7} is the final conclusion.
\section{Background} 
\label{basic_QSVM}

\subsection{SVM algorithm}
The SVM classification algorithm can classify data into two sub-groups. Let us assume there are $M$ training data points $\vec{x_i}, i=1, \ldots, M$, and each data has a label $+1$ or $-1$. Let us also assume that the number of dimensions of the feature space is $N$. In this case, the training data can be written as $\{(\vec{x_i}, y_i): \vec{x_i} \in \mathbb{R}^N, y_i=\pm 1\}$.

\begin{figure}[h!]
  \centering
\includegraphics[width=6cm]{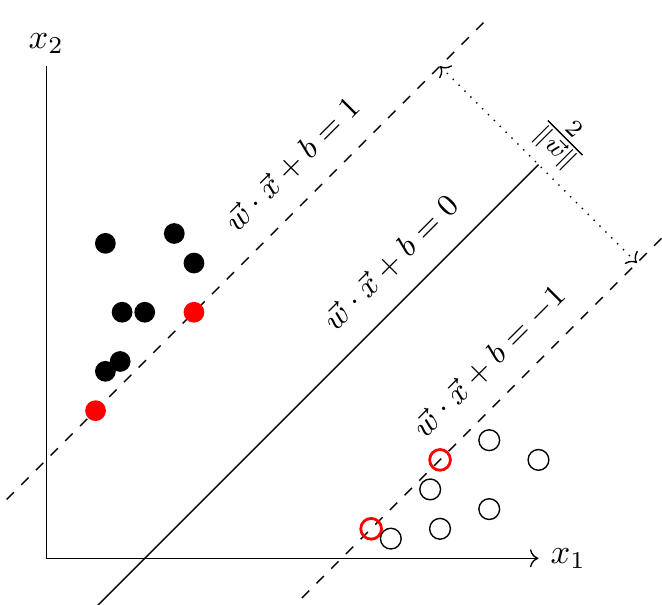}
 \caption{An example of SVM classification}
  \label{SVMexample}
  \end{figure}

Fig. \ref{SVMexample} shows an example of SVM classification with $N=2$. As shown in the figure, the goal of SVM is to find the maximum-margin hyperplane $\vec{w}\cdot\vec{x}+b=0$ that can divide the data into two classes and make the distance between the classes the furthest. The number of dimensions of the hyperplane is $N-1$, which in our case equals to $1$. 

The training data points $\vec{x_i} $ will be classified into the $+1$ class if the inequality $\vec{w}\cdot\vec{x_i}+b\geqslant 1$ is fulfilled; or the data points will be classified into the $-1$ class if the inequality $\vec{w}\cdot\vec{x_i}+b\leqslant -1$ is fulfilled. Thus, the classification set of equations can be written as:

\begingroup
\fontsize{8pt}{8pt}\selectfont
\begin{equation}
    \left\{\begin{matrix}
y_i=+1\textrm{ if }\vec{w}\cdot\vec{x_i}+b\geqslant +1\\ 
y_i=-1\textrm{ if }\vec{w}\cdot\vec{x_i}+b\leqslant -1
\end{matrix}\right.
\end{equation}
\endgroup

or in a simpler manner:

\begingroup
\fontsize{8pt}{8pt}\selectfont
\begin{equation}
y_i(\vec{w}\cdot\vec{x_i}+b)\geqslant 1
\label{SVM_constraint}
\end{equation}
\endgroup

The margin between two classes can also be represented by the distance between two support hyperplanes $\vec{w}\cdot\vec{x}+b=1$ and $\vec{w}\cdot\vec{x}+b= -1$. With the expressions of the support hyperplanes defined, the distance between them can be represented by $\frac{2}{\left \| \vec{w} \right \|}$. Thus,  SVM aims at maximizing the margin $\frac{2}{\left \| \vec{w} \right \|}$, which is the same as minimizing $\frac{\left \| \vec{w} \right \|^2}{2}$. The Lagrange's method is then used to solve this problem \cite{boser1992training}.

\subsection{Noisy Intermediate Scale Quantum (NISQ) Computers}

Noisy Intermediate Scale Quantum (NISQ) computers refer to near-term $50-100$ qubits quantum computers that can solve some problems that classical computers cannot, but they are not advanced enough to realize fully fault-tolerant quantum computation \cite{preskill2018quantum}. NISQ computers do not have enough qubits to do the error correction, so they have to directly use the imperfect physical qubits to implement quantum computing algorithms and operations.

Thus, when we run quantum circuits on real quantum computers, we have to consider their imperfectness, such as the decoherence. Quantum decoherence, i.e., the loss of quantum coherence, describes the phenomenon that qubits in a quantum computer will lose their quantum mechanical properties as the time goes by due to their interaction with the environment \cite{zurek2003decoherence}.  Because of the limited coherence time of quantum computers, only quantum circuits that are short enough can be reliably run on them, and there is a need to reduce the circuit depth, which is the length of the longest path in a quantum circuit. Otherwise, the result will not be as expected.

\subsection{Basic Quantum Computing Concepts}
\subsubsection{One single qubit and quantum superposition}
The bit is the fundamental unit of a classical computer and has two possible states: $0$ and $1$. Similarly, the qubit (quantum bit) is the fundamental unit of a quantum computer, and it also has two possible states $0$ and $1$ but only after measurement or observation. Before measurement, a qubit can stay in both $0$ and $1$ simultaneously, i.e., superposition.

A qubit can be written as a unit vector in a two-dimensional complex vector space $\mathbb{C}^2$. A qubit in state zero is written as $|0\rangle$; a qubit in state one is written as $|1\rangle$, and they are defined as:

\begingroup
\fontsize{8pt}{8pt}\selectfont
\begin{equation}
\begin{aligned}
    |0\rangle=\begin{pmatrix}
    1\\ 
    0
    \end{pmatrix}
    \\
    |1\rangle=\begin{pmatrix}
    0\\ 
    1
    \end{pmatrix}
\end{aligned}   
\end{equation}
\endgroup

where $|\cdot\rangle$ is a ket, and it is a standard notation for describing quantum states. Ket notation together with its conjugate transpose, the bra $\langle\cdot|$, is known as the Dirac notation.

\begin{figure}[h!]
    \centering
    \begin{tikzpicture}[line cap=round, line join=round, >=Triangle]
      \clip(-2.19,-2.49) rectangle (2.66,2.58);
      \draw [shift={(0,0)}, lightgray, fill, fill opacity=0.1] (0,0) -- (56.7:0.4) arc (56.7:90.:0.4) -- cycle;
      \draw [shift={(0,0)}, lightgray, fill, fill opacity=0.1] (0,0) -- (-135.7:0.4) arc (-135.7:-33.2:0.4) -- cycle;
      \draw(0,0) circle (2cm);
      \draw [rotate around={0.:(0.,0.)},dash pattern=on 3pt off 3pt] (0,0) ellipse (2cm and 0.9cm);
      \draw (0,0)-- (0.70,1.07);
      \draw [->] (0,0) -- (0,2);
      \draw [->] (0,0) -- (-0.81,-0.79);
      \draw [->] (0,0) -- (2,0);
      \draw [dotted] (0.7,1)-- (0.7,-0.46);
      \draw [dotted] (0,0)-- (0.7,-0.46);
      \draw (-0.08,-0.3) node[anchor=north west] {$\varphi$};
      \draw (0.01,0.9) node[anchor=north west] {$\theta$};
      \draw (-1.01,-0.72) node[anchor=north west] {$\mathbf {x}$};
      \draw (2.07,0.3) node[anchor=north west] {$\mathbf {y}$};
      \draw (-0.5,2.6) node[anchor=north west] {$\mathbf {z}=|0\rangle$};
      \draw (-0.4,-2) node[anchor=north west] {$-\mathbf {z}=|1\rangle$};
      \draw (0.4,1.65) node[anchor=north west] {$|\psi\rangle$};
      \scriptsize
      \draw [fill] (0,0) circle (1.5pt);
      \draw [fill] (0.7,1.1) circle (0.5pt);
    \end{tikzpicture}
    \caption{Bloch sphere representation of a qubit}
    \label{fig:bloch sphere}
\end{figure}
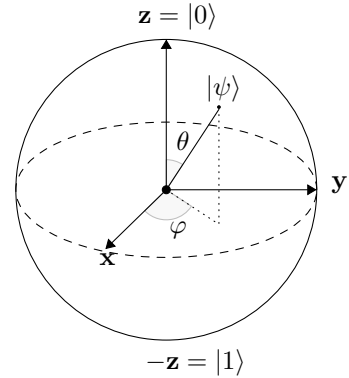

Any arbitrary one-qubit quantum state $|\psi\rangle$ can be represented by:

\begingroup
\fontsize{8pt}{8pt}\selectfont
\begin{equation}
    |\psi\rangle=a_0|0\rangle+a_1|1\rangle=a_0\begin{pmatrix}
    1\\ 
    0
    \end{pmatrix}+a_1\begin{pmatrix}
    0\\ 
    1
    \end{pmatrix}= \begin{pmatrix}
    a_0\\ 
    a_1
    \end{pmatrix}
\end{equation}
\endgroup

where $|a_0|^2+|a_1|^2=1$. This shows one of the most important characteristics of qubit, superposition, which means that a qubit can stay in state zero and state one at the same time until it is observed. According to the quantum mechanics laws, after measurement or observation, the qubit will collapse in one or the other state and there will be a probability $|a_0|^2$ to observe $|0\rangle$ and a probability $|a_1|^2$ to observe $|1\rangle$. The one-qubit quantum state $|\psi\rangle$ can also be regarded as a point $(\theta, \phi)$ on the Bloch sphere, as shown in Fig. \ref{fig:bloch sphere}, where $a_0=cos(\frac{\theta}{2})$ and $a_1=e^{i\phi}sin(\frac{\theta}{2})$. 

For example, a qubit can be in the superposition state $\frac{1}{\sqrt{2}}(|0\rangle+|1\rangle)$, which can be generated through a Hadamard gate on a $|0\rangle$ quantum state. After measuring this qubit, the probabilities of getting zero or one states are both $\frac{1}{2}$.

\subsubsection{Multiple qubits and quantum entanglement}
Qubits can also be combined together to form multi-qubit system, whose matrix form can be calculated by the tensor product of all qubits. For example, a two-qubit system can be represented by four basis states: $|00\rangle$, $|01\rangle$, $|10\rangle$, $|11\rangle$, and each of them, in its matrix form, is a four-dimensional vector:

\begingroup
\fontsize{8pt}{8pt}\selectfont
\begin{equation}
    |00\rangle=|0\rangle\otimes|0\rangle=\begin{pmatrix}
1\\ 
0
\end{pmatrix}\otimes\begin{pmatrix}
1\\ 
0
\end{pmatrix}=\begin{pmatrix}
1\otimes\begin{pmatrix}
1\\ 
0
\end{pmatrix}\\ 
0\otimes\begin{pmatrix}
1\\ 
0
\end{pmatrix}
\end{pmatrix}=\begin{pmatrix}
1\\ 
0\\ 
0\\ 
0
\end{pmatrix}
\end{equation}
\endgroup

Note that the other three basis states can be obtained similarly. A two-qubit system can exist in a superposition of these four basis states:

\begingroup
\fontsize{8pt}{8pt}\selectfont
\begin{align}
\allowdisplaybreaks
   |\psi\rangle&=a_{0}|00\rangle+a_{1}|01\rangle+a_{2}|10\rangle+a_{3}|11\rangle \notag \\
&= a_{0}\begin{pmatrix}
1\\ 
0\\ 
0\\ 
0
\end{pmatrix}+a_{1}\begin{pmatrix}
0\\ 
1\\ 
0\\ 
0
\end{pmatrix}+a_{2}\begin{pmatrix}
0\\ 
0\\ 
1\\ 
0
\end{pmatrix}+a_{3}\begin{pmatrix}
0\\ 
0\\ 
0\\ 
1
\end{pmatrix}=\begin{pmatrix}
a_{0}\\ 
a_{1}\\ 
a_{2}\\ 
a_{3}
\end{pmatrix}
\end{align}
\endgroup

where $a_{0}$ to $a_{3}$ are the amplitudes (also named weights and coefficients) of the basis. Similar to the case of one qubit, $|a_{i}|^2$ represents the probability of measuring state $|i\rangle$, and the sum of $|a_{i}|^2$ for all states shall be 1:

\begingroup
\fontsize{8pt}{8pt}\selectfont
\begin{equation}
    \sum_{i=0}^{3}|a_{i}|^2=1
\end{equation}
\endgroup

Generalizing, every $n$-qubit quantum state can be represented by a $2^n$-dimensional complex vector in the Hilbert space:

\begingroup
\fontsize{8pt}{8pt}\selectfont
\begin{equation}
    |\psi\rangle=\sum_{i=0}^{2^n-1}a_{i}|i\rangle
\end{equation}
\endgroup

After the measurement, the probability of $|\psi\rangle$ to collapse into state $|i\rangle$ is $|a_{i}|^2$ where $\sum_{i=0}^{2^n-1}|a_{i}|^2=1$.

Besides superposition, the entanglement among multiple qubits is another important characteristic of qubits. Entanglement is a quantum physics phenomenon, which describes the interaction of pairs or groups of qubits. The state of each qubit cannot be described independently; rather, it has to be regarded as a whole. 

\begin{figure}[h!] $\;$
  \centering
  \Qcircuit @C=1em @R=.7em {
\lstick{|0\rangle}&\gate{H} & \ctrl{1} & \qw \\
\lstick{|0\rangle}& \qw     & \targ    & \qw 
}
  \caption{The generation of the Bell state}
  \label{Bell}
\end{figure}
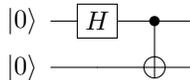

Examples for quantum entanglement are the Bell states, also called the Einstein-Podolsky-Rosen (EPR) states. One of the Bell states is $\frac{1}{\sqrt{2}}(|00\rangle+|11\rangle)$, which indicates two entangled qubits in two possible quantum states, $|00\rangle$ and $|11\rangle$, such that when the first qubit is in the zero state, the other one is also in the zero state; when the first qubit is in the one state, the other one is also in the one state. The two qubits cannot be described independently, but we can describe them as a whole: they are always in the same state. This Bell state $\frac{1}{\sqrt{2}}(|00\rangle+|11\rangle)$ is also denoted as $|+\rangle$, and it is generated through two quantum gates, the Hadamard gate and the CNOT (controlled NOT) gate, as shown in Fig. \ref{Bell}.

The number of basis states a quantum system contains increases exponentially with the number of qubits. 
Thus, the properties of superposition and entanglement
can enable one qubit to contain two bits of information; two qubits to contain four bits of information; $n$ qubits to contain $2^n$ bits of information. Therefore, quantum computers have a much larger computational power than classical computers.

\subsubsection{Quantum gate and quantum circuit}
In classical computers, there are logic gates, such as AND gate, OR gate, XOR gate, and NOT gate, etc., and these are the building blocks of conventional digital circuits. In quantum computers, we have quantum logic gates and can transform quantum states to other quantum states. Table \ref{gatetable} relates some fundamental quantum gates and their matrix representation.

\begin{table}
\centering
  \includegraphics[width=8.5cm]{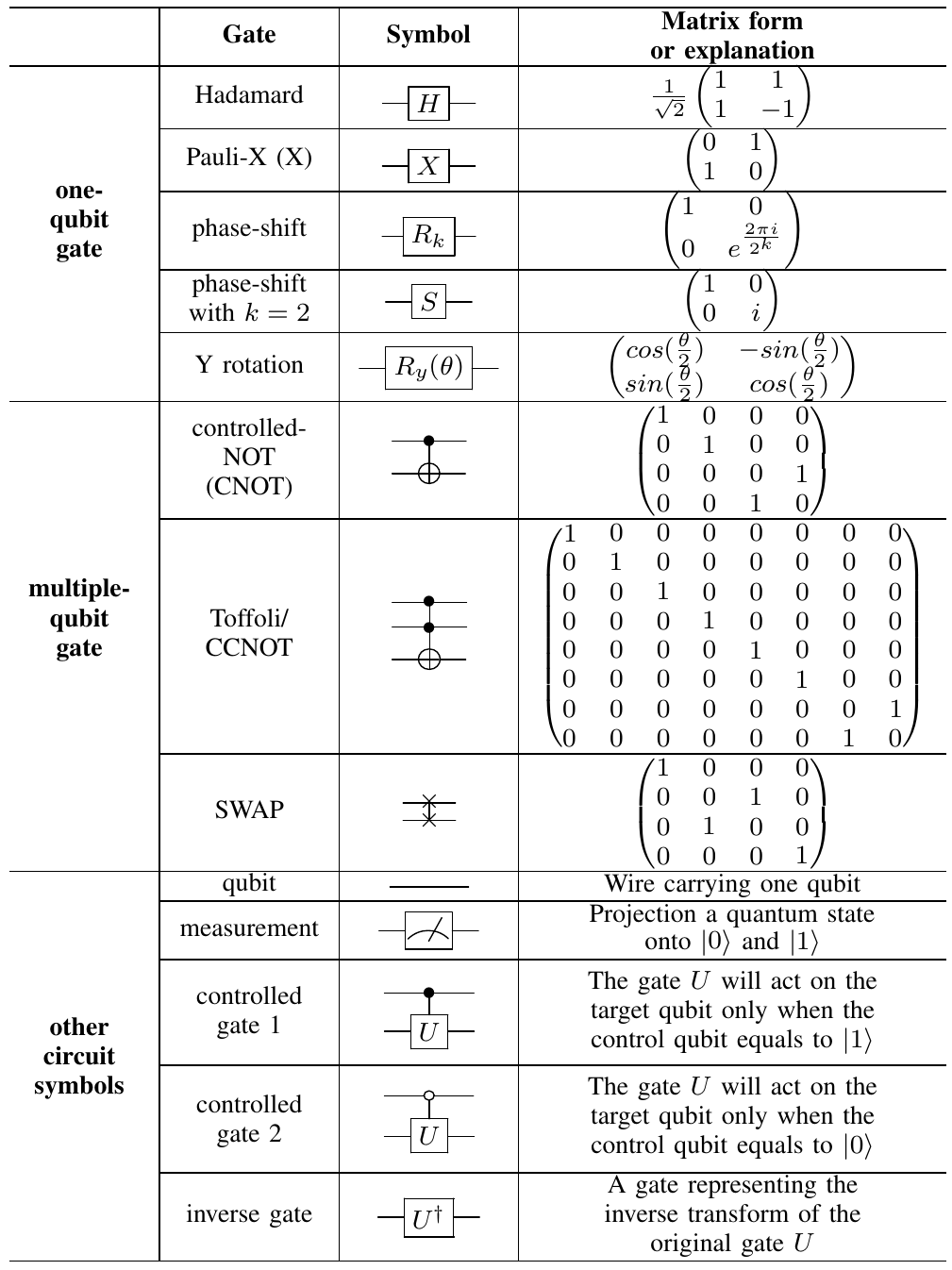}
  \caption{Quantum gates and quantum circuit symbols}
  \label{gatetable}
\end{table}

Different from the classical logical gates, there exist infinite different quantum gates. This is because quantum gates can be represented by unitary matrices and any transform using the unitary matrix can be considered a quantum gate. An $N\times N$ complex matrix $U$ is a unitary matrix if it satisfies:

\begingroup
\fontsize{8pt}{8pt}\selectfont
\begin{equation}
    U^\dagger U=UU^\dagger=I_N
\end{equation}
\endgroup

where $U^\dagger$ is the conjugate transpose of $U$, and $I_N$ is the $N\times N$ identity matrix.  

The number of input qubits and output qubits of the quantum gate should be the same, and a quantum gate acting on $n$ qubits should be represented by a $2^{n}\times 2^{n}$ unitary matrix. Thus, a quantum gate operating on a quantum state can be calculated by a point multiplication of a $2^{n}\times 2^{n}$ matrix and a $2^{n}\times 1$ vector. The result of the transform is a new quantum state represented by a vector of the same size as the original vector.

A series of quantum gates in a designed sequence will build up to a quantum circuit. A quantum circuit is designed to transform an initial quantum state into a quantum state with our desired probability distribution. 

\section{Related Work}
\subsection{QSVM algorithm}
There are two methods for implementing QSVM on a quantum computer. One is based on Grover's Search algorithm, with quadratic speedup \cite{anguita2003quantum}; the other one is based on the HHL algorithm, with exponential speedup \cite{rebentrost2014quantum}. The HHL algorithm \cite{harrow2009quantum}, named after its authors, Aram Harrow, Avinatan Hassidim, and Seth Lloyd, is the quantum algorithm for linear systems of equations. It can extract certain properties of $\vec{x}$ satisfying $A\vec{x}=\vec{b}$, where $A$ is an $N\times N$ matrix, and $\vec{b}$ is a vector with size $N\times 1$. The computational complexity of the classical SVM algorithm is $O[log(\epsilon^{-1})poly(NM)]$, and it is proportional to the polynomial in $NM$. Here $N$ is the number of dimensions of the data, $M$ is the number of training data, and $\epsilon$ is the accuracy. The QSVM based on the HHL algorithm, as in ref. \cite{rebentrost2014quantum}, can achieve $O[log(NM)]$ performance on both training and testing processes, and thus, it can exponentially speed up a calculation when compared to the classical SVM algorithm.

As shown in \cite{rebentrost2014quantum}, by employing the least-squares reformulation of the support vector machine described in \cite{suykens1999least}, we can change the original SVM problem, a quadratic programming problem, into a problem of solving a linear equation system:

\begingroup
\fontsize{8pt}{8pt}\selectfont
\begin{equation}
    F\begin{pmatrix}
b\\ 
\vec{\alpha}
\end{pmatrix}\equiv\begin{pmatrix}
0 & \vec{1}^T\\ 
 \vec{1}&  K+\gamma^{-1}I
\end{pmatrix}\begin{pmatrix}
b\\ 
\vec{\alpha}
\end{pmatrix}=\begin{pmatrix}
0\\ 
\vec{y}
\end{pmatrix} \label{svmequation}
\end{equation}
\endgroup

where $K$ is the $M\times M$ kernel matrix and its elements can be calculated by $K_{jk}=k(x_j, x_k)=\vec{x_j}\cdot \vec{x_k}$ when a linear kernel is chosen; $\gamma$ is a user-defined value to control the trade-off between training error and SVM objective; $\vec{y}$ is a vector storing the labels of the training data, and $I$ is the unit matrix. So the only unknown term in this linear equation is the vector $\begin{pmatrix}
b\\ 
\vec{\alpha}
\end{pmatrix}$. In here, both $\vec{\alpha}$ and $b$ are parameters for calculating the SVM classifier, the decision hyperplane for splitting data into two sub-groups. Once the parameters of the hyperplane are determined, thanks to the linear system solving algorithm, the HHL algorithm, a new data point $\vec{x_0}$ can be classified according to:

\begingroup
\fontsize{8pt}{8pt}\selectfont
\begin{align}
   y(\vec{x_0})&= sgn(\vec{w}\cdot\vec{x_0}+b)\notag 
   \\&=sgn(\sum_{i=1}^{M}\alpha_ik(\vec{x_i}\cdot\vec{x_0})+b)
    \label{Classification-formula}
\end{align}
\endgroup

where $\vec{x_i}$ with $i=1, \ldots, M$ is the training data; $\alpha_i$ is the $i^{th}$ dimension of the parameter $\vec{\alpha}$; $\vec{w}$ is the slope of the hyperplane, which can be calculated through the parameter $\vec{\alpha}$. Parameter $b$ is the offset of the hyperplane, which in our case is chosen to be $0$, and the sign function $sgn()$ is defined as:

\begingroup
\fontsize{8pt}{8pt}\selectfont
\begin{equation}
    sgn(x)=\left\{\begin{matrix}
1,\;\textrm{if} \;x \geqslant 0\\ 
-1,\; \textrm{if} \;x<0
\end{matrix}\right.
\end{equation}
\endgroup

\subsection{QSVM implementation}
The authors in ~\cite{li2015anexperimental} introduce a method to implement the QSVM with exponential speedup based on the physical implementation of the Nuclear Magnetic Resonance (NMR) technique. The QSVM system in this implementation can be used in the optical character recognition (OCR) problem to distinguish between the handwritten ``6" and ``9" images. 
In this reference, the problem is further simplified by using the non-offset reduction:  $b$ is set to $0$, so Equation~\ref{svmequation} can be re-written as:

\begingroup
\fontsize{8pt}{8pt}\selectfont
\begin{equation}
    F\vec{\alpha}=\vec{y}
\end{equation}
\endgroup

The implementation process described in \cite{li2015anexperimental} can be divided into three parts: 1) the preprocessing unit, 2) the generation of kernel matrix $K$, and 3) the QSVM quantum circuit for classification and the result readout. However, there exist some drawbacks in this implementation:

\begin{enumerate}
    \item The QSVM system is limited to classify the handwritten ``6" and ``9" images in the specific OCR dataset, but is not designed for other datasets.
    \item The $arccot()$ function in the preprocessing unit to calculate the rotations is only suitable for data in certain quadrants, which will limit the classification accuracy.
    \item When generating the kernel matrix, the time-consuming quantum tomography \cite{nielsen2010quantum} is needed to calculate the density matrix of the first qubit of the circuit.
    \item The quantum circuit for QSVM classification is too complex to be executed on a superconducting quantum computer, for example, the IBMQX2 quantum computer provided by IBM.
\end{enumerate}

\section{Our implementation of QSVM system for NISQ computers}
In this section, we propose an implementation of QSVM system for NISQ computers, which can overcome the problems of reference \cite{li2015anexperimental}. The implementation consists of three units: the preprocessing unit, the generation of kernel matrix and the optimization of the HHL quantum circuit for QSVM classification. 
\subsection{Preprocessing unit}
\begin{figure}[h!]
  \centering
  \includegraphics[width=9cm]{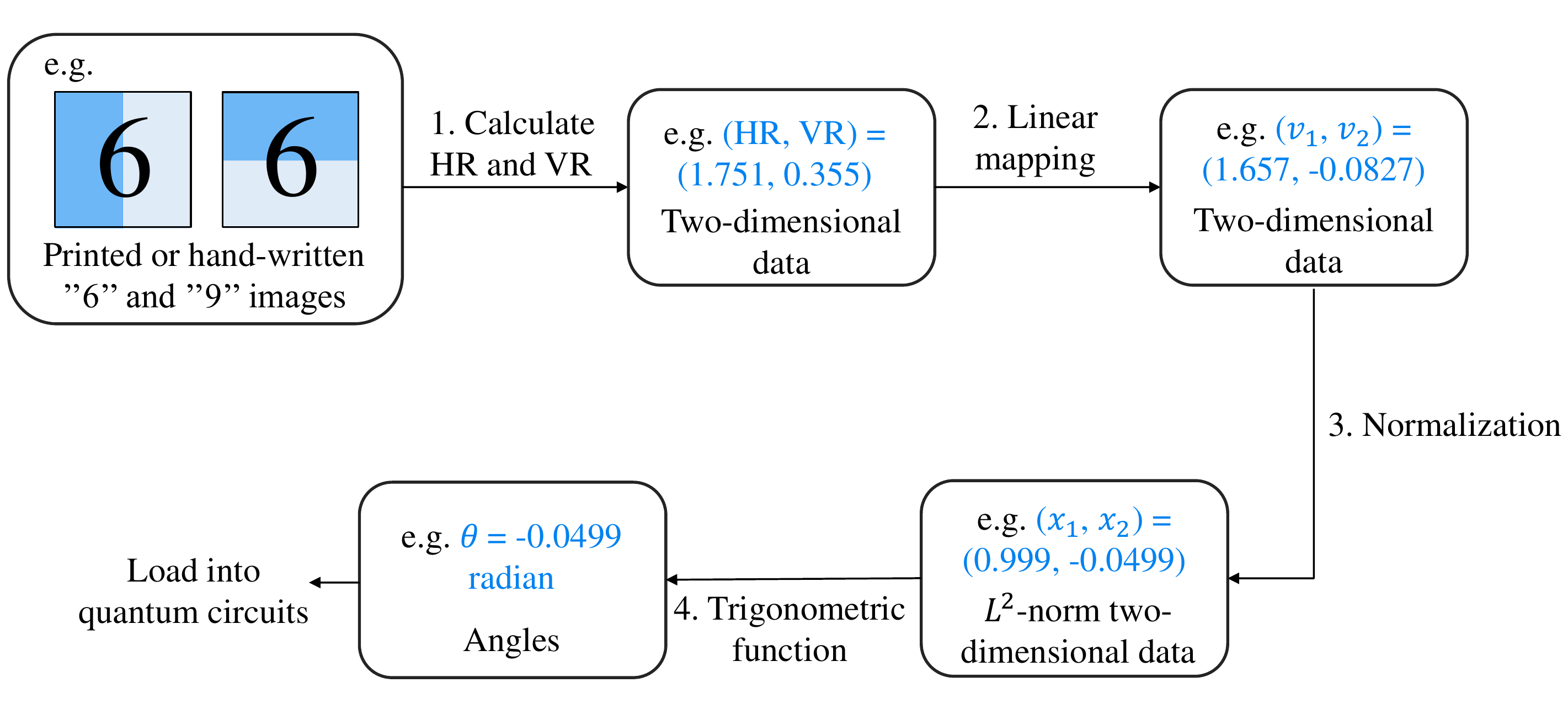}
  \caption{Data preprocessing flow chart}
  \label{preprocessing_flow}
\end{figure}

As shown in Fig. \ref{preprocessing_flow}, the preprocessing unit comprises four sub-steps: i) calculating the horizontal (HR) and the vertical ratios (VR); ii) linear mapping; iii) $L^2$ normalization and, iv) calculation of the rotation angles. 
\subsubsection{Step i. Calculating the HR and VR}
This step is specific for the OCR dataset, and it  converts the character images data into two-dimensional data $(\textrm{HR}_i, \textrm{VR}_i)$, where $i=1, \ldots, m$ and $m$ is the number of test data. The ratios are defined as:

\begingroup
\fontsize{8pt}{8pt}\selectfont
\begin{equation}
\begin{aligned}
    \textrm{HR}=\frac{\textrm{number of black pixels in the \textbf{left} half}}{\textrm{number of black pixels in the \textbf{right} half}}
    \\
    \textrm{VR}=\frac{\textrm{number of black pixels in the \textbf{upper} half}}{\textrm{number of black pixels in the \textbf{lower} half}}
\end{aligned}
\label{HRVR}
\end{equation}
\endgroup

\subsubsection{Step ii. Linear mapping}
In this step, we use the linear mapping to adapt the training data and test data to make the matrix $F=K+\gamma^{-1}I$ generated by the training data to have a simpler form. For the OCR dataset, the linear mapping is given by
\cite{li2015anexperimental}:

\begingroup
\fontsize{8pt}{8pt}\selectfont
\begin{equation}
\begin{aligned}
    (\vec{v_i})_1=\textrm{HR}_i\times1.3-0.62
    \\
    (\vec{v_i})_2=\textrm{VR}_i\times0.95-0.42
\end{aligned}
\label{equ:linear mapping}
\end{equation}
\endgroup

To make the QSVM system suitable for classifying not only the OCR dataset but any other two-dimensional datasets that can be linearly separated by a line crossing the origin, we need to modify the previous linear mapping. In this step, a way of calculating the new coefficients for the linear mapping is provided, so no matter how the dataset changes, we can always use the same QSVM quantum circuit, which has the same decision model to classify the test data.

The linear mapping maps both the training and test data. We map the training data to ensure that the eigenvalues of the matrix $F$ can be represented by two qubits, and we need to do the same thing to the test data as well. 
Let us consider an example without linear mapping. In this case, the matrix $F$ calculated from the training data may have the following values: $\begin{pmatrix}
 1& 0.8\\ 
 0.8& 1
\end{pmatrix}$. The calculated eigenvalues of this matrix are $0.2$ and $1.8$, which can be denoted by the binary numbers $0.0011\ldots$
and $1.1100\ldots$, and to store these binary eigenvalues, we need a lot of qubits.  To reduce the length of eigenvalues and the number of qubits, we can linearly map the training data to make $F=\begin{pmatrix}
 1& 0.5\\ 
 0.5& 1
\end{pmatrix}$. Now, the eigenvalues are $0.5$ (binary representation is $0.1$) and $1.5$ (binary representation is $1.1$), and can be represented by two qubits.

Thus, for datasets other than the OCR dataset, we just need to recalculate the coefficients in Equation \ref{equ:linear mapping}. The new equations are then defined as follow: 

\begingroup
\fontsize{8pt}{8pt}\selectfont
\begin{equation}
\begin{aligned}
    (\vec{v_i})_1=(t_{i})_1\times a+b
    \\
    (\vec{v_i})_2=(t_{i})_2\times c+d
\end{aligned}
\label{new_linear_mapping}
\end{equation}
\endgroup

where $(t_{i})_1$ and $(t_{i})_2$ are the first and second dimensions of the training data of a new dataset. After this new linear mapping and the $L^2$ normalization, the training data  $((\vec{x_i})_1,(\vec{x_i})_2)=(\frac{(\vec{v_i})_1}{|\vec{v_i}|},\frac{(\vec{v_i})_2}{|\vec{v_i}|})$ will equal to the OCR dataset's preprocessed training data, which are $(0.987, 0.159)$ and $(0.345, 0.935)$, and the kernel matrix $K$ up to a constant factor $tr(K)$, $\hat{K}=K/tr(K)$, can be approximated to $\begin{pmatrix}
 0.5& 0.25\\ 
 0.25& 0.5
\end{pmatrix}$. From Equation \ref{equ:linear mapping} and knowing the values of the training data $n_1,n_2,n_3$ and $n_4$, then we can solve the unknown paramenters $a$, $b$, $c$, $d$ by the solving the equations:

\begingroup
\fontsize{8pt}{8pt}\selectfont
\begin{equation}
    \frac{a(t_1)_1+b}{\sqrt{(a(t_1)_1+b)^2+(c(t_1)_2+d)^2}}=0.987\equiv n_1
    \label{fangcheng_1}
\end{equation}
\begin{equation}
    \frac{c(t_1)_2+d}{\sqrt{(a(t_1)_1+b)^2+(c(t_1)_2+d)^2}}=0.159\equiv n_2
    \label{fangcheng_2}
\end{equation}
\begin{equation}
    \frac{a(t_2)_1+b}{\sqrt{(a(t_2)_1+b)^2+(c(t_2)_2+d)^2}}=0.345\equiv n_3
\end{equation}
\begin{equation}
    \frac{c(t_2)_2+d}{\sqrt{(a(t_2)_1+b)^2+(c(t_2)_2+d)^2}}=0.935\equiv n_4
    \label{fangcheng_4}
\end{equation}
\endgroup

where $(t_1)_1$, $(t_1)_2$, $(t_2)_1$ and $(t_2)_2$ are the first and second dimensions of the two training data of the new dataset and are known values. The analytic equations for $a$, $b$, when $c$ and $d$ take any arbitrary values, are:

\begingroup
\fontsize{8pt}{8pt}\selectfont
\begin{equation}
\begin{aligned}
    a= \frac{n_1\cdot n_4 \cdot(c(t_1)_2+d)-n_2\cdot n_3\cdot(c(t_2)_2+d)}{n_2\cdot n_4\cdot((t_1)_1-(t_2)_1)} 
    \\
    b=\frac{x_1 \cdot n_2 \cdot n_3(c(t_2)_2+d)-x_2 \cdot n_1 \cdot n_4(c(t_1)_2+d)}{n_2\cdot n_4\cdot((t_1)_1-(t_2)_1)}
\label{final_formula}
\end{aligned}
\end{equation}
\endgroup

Then we can input the calculated coefficients $a$ and $b$, and the arbitrary chosen coefficients $c$ and $d$ to the left-hand side of Equation~\ref{fangcheng_1} and~\ref{fangcheng_2}, to check whether the $n_1$ and $n_2$ are positive. The $n_1$ and $n_2$ might be negative numbers, because we introduced the square and square root operations into the previous equations. If $n_1$ and $n_2$ are negative, we can choose the values of $c$ and $d$ again so that $n_1$ and $n_2$ become positive. That will give us the proper values for $a$, $b$, $c$ and $d$.

\subsubsection{Step iii. $L^2$ normalization}
This step transfers the two-dimensional data $((\vec{v_i})_1, (\vec{v_i})_2)$ into $((\vec{x_i})_1, (\vec{x_i})_2)$, fulfilling $|(\vec{x_i})_1|^2+|(\vec{x_i})_2|^2=1$, so each data $((\vec{v_i})_1, (\vec{v_i})_2)$ is normalized to a point on the unit circle. The normalization is defined as:

\begingroup
\fontsize{8pt}{8pt}\selectfont
\begin{equation}
\begin{aligned}
    (\vec{x_i})_1=\frac{(\vec{v_i})_1}{|\vec{v_i}|}
    \\
    (\vec{x_i})_2=\frac{(\vec{v_i})_2}{|\vec{v_i}|}
\end{aligned}
\end{equation}
\endgroup

where each data point then can be denoted by a one-qubit quantum state:

\begingroup
\fontsize{8pt}{8pt}\selectfont
\begin{equation}
    |x_i\rangle = (\vec{x_i})_1|0\rangle+(\vec{x_i})_2|1\rangle
\end{equation}
\endgroup

\subsubsection{Step iv. Calculation of the rotation angles}
The one-qubit quantum state $|x_i\rangle $ is usually obtained by a rotation gate acting on an initial quantum state $|0\rangle $, with a certain angle. The authors in reference~\cite{li2015anexperimental} use the $arccot()$ function to calculate the rotation angle, which is only suitable for test data in specific quadrants. 

\begin{figure}[h!]
  \centering
  \includegraphics[width=8.7cm]{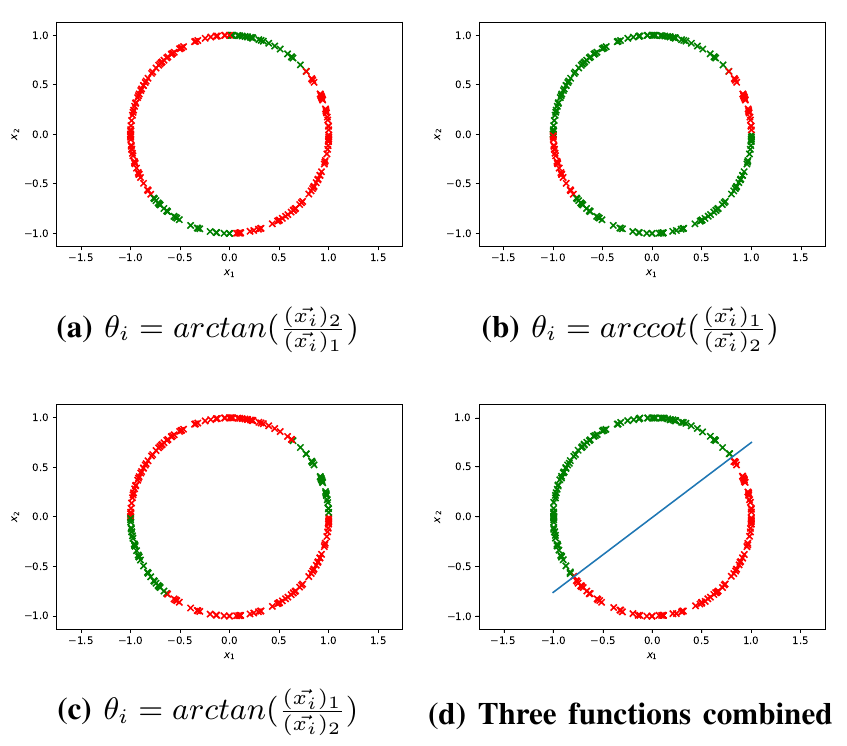}
  \caption{Choosing trigonometric functions for calculating angles}
  \label{trigonometric functions}
\end{figure}

The results of classifying $100$ random-generated points by calculating the angles with $\theta_i=arccot(\frac{(\vec{x_i})_1}{(\vec{x_i})_2})$ are shown in Fig.~\ref{trigonometric functions}b. As shown from the graph, the points are not only classified by the expected boundary, but also by the x-axis. Thus,  the classification is correct only if the data points are in the first or second quadrant. If, instead, we calculate the angles $\theta_i$ by $\theta_i=arctan(\frac{(\vec{x_i})_2}{(\vec{x_i})_1})$, the classification results for $100$ random-generated points on the unit circuit are shown in Fig.~\ref{trigonometric functions}a. Note that the points are not only classified by the expected boundary, but also by the y-axis. Thus,  the classification is correct only if the data points are in the first or fourth quadrant. Similarly, if we calculate the angles by the trigonometric function $\theta_i=arctan(\frac{(\vec{x_i})_1}{(\vec{x_i})_2})$, the classification results of the $100$ data are shown in Fig.~\ref{trigonometric functions}c. 

To make the classification results correct regardless of the quadrant the data points belong to, we combine the above three functions to calculate the  angles. If $((\vec{x_i})_1,(\vec{x_i})_2)$ are known, then the set of classifying equations are: 

\begingroup
\fontsize{8pt}{8pt}\selectfont
\begin{equation}
\left\{\begin{matrix}
1^{st} \textrm{ quadrant, }\theta_i=arctan(\frac{(\vec{x_i})_2}{(\vec{x_i})_1}) \textrm{ or }arccot(\frac{(\vec{x_i})_1}{(\vec{x_i})_2})\\ 
2^{nd}\textrm{ quadrant, }\theta_i=arccot(\frac{(\vec{x_i})_1}{(\vec{x_i})_2})\;\;\;\;\;\;\;\;\;\;\;\;\;\;\;\;\;\;\;\;\;\;\;\;\;\;\; \\ 
3^{rd}\textrm{ quadrant, }\theta_i=arctan(\frac{(\vec{x_i})_1}{(\vec{x_i})_2})\;\;\;\;\;\;\;\;\;\;\;\;\;\;\;\;\;\;\;\;\;\;\;\;\;\;\\\ 
4^{th}\textrm{ quadrant, }\theta_i=arctan(\frac{(\vec{x_i})_2}{(\vec{x_i})_1})\textrm{ or }arctan(\frac{(\vec{x_i})_1}{(\vec{x_i})_2})
\end{matrix}\right.
\label{combineequation}
\end{equation}
\endgroup

The results after using the equations above  are shown in Fig.~\ref{trigonometric functions}d. The points are now only separated by the classification boundary, as we expected.


\subsection{Generation of kernel matrix}
\subsubsection{Training-data oracle without quantum tomography}
Reference \cite{li2015anexperimental} provides a quantum circuit with a training-data oracle to generate the kernel matrix, followed by the quantum tomography technique to calculate the density matrix of the first qubit. Quantum tomography, or more specifically the quantum state tomography, reconstructs the quantum state or density matrix of a circuit by measuring different observables on many identically fabricated circuit replicas, so it is a time-consuming technique that needs a lot of observations \cite{riofrio2011continuous}. We propose to use a classical method for result readout of the training-data oracle circuit in \cite{li2015anexperimental}, which only uses the raw counts of the possible states of the circuit, and does not need the quantum state tomography. The method is the following:

After measuring the quantum state of the training-data oracle, we get a 2-qubit state such as $|\psi\rangle=\alpha_0|00\rangle+\alpha_1|01\rangle+\alpha_2|10\rangle+\alpha_3|11\rangle$, where the amplitudes $\alpha_0$ to $\alpha_3$ can be obtained by the raw counts of the possible states. We can first further calculate $|\psi\rangle$ as: 

\begingroup
\fontsize{8pt}{8pt}\selectfont
\begin{align}
   |\psi\rangle&= a_0|00\rangle+a_1|01\rangle+a_2|10\rangle+a_3|11\rangle \notag \\&=(|a_0|^2+|a_2|^2)\begin{pmatrix}
\bar{a_0}\\ 
\bar{a_2}
\end{pmatrix}\otimes|0\rangle+(|a_1|^2+|a_3|^2)\begin{pmatrix}
\bar{a_1}\\ 
\bar{a_3}
\end{pmatrix}\otimes|1\rangle \notag\\
             &\overset{def}{=}p_0 |\psi_0\rangle\otimes|0\rangle +p_1 |\psi_1\rangle\otimes|1\rangle
\end{align}
\endgroup

where $\bar{a_0}$, $\bar{a_1}$, $\bar{a_2}$, $\bar{a_3}$ are the normalized amplitudes, which satisfy:

\begingroup
\fontsize{8pt}{8pt}\selectfont
\begin{equation}
\begin{aligned}
    |\bar{a_0}|^2+|\bar{a_2}|^2=1\\
    |\bar{a_1}|^2+|\bar{a_3}|^2=1
\end{aligned}
\end{equation}
\endgroup

Then the density matrix $\rho$ of the first qubit can be calculated by:

\begingroup
\fontsize{8pt}{8pt}\selectfont
\begin{align}
\rho&=p_0|\psi_0 \rangle \langle \psi_0 |+p_1|\psi_1 \rangle \langle \psi_1 |\notag\\
             &= p_0 \begin{pmatrix}
\bar{a_0}\\ 
\bar{a_2}
\end{pmatrix}\begin{pmatrix}
 \bar{a_0}& \bar{a_2}
\end{pmatrix}+p_1 \begin{pmatrix}
\bar{a_1}\\ 
\bar{a_3}
\end{pmatrix}\begin{pmatrix}
 \bar{a_1}& \bar{a_3}
\end{pmatrix}
\end{align}
\endgroup

And this density matrix $\rho$ is equal to $\hat{K}=K/tr(K)$ \cite{li2015anexperimental}, which is the kernel matrix divided by its trace $tr(K)$, calculated by the sum of the main diagonal elements. 
\\
\subsubsection{Trade-off between the circuit depth and the number of qubits in  the training-data oracle }
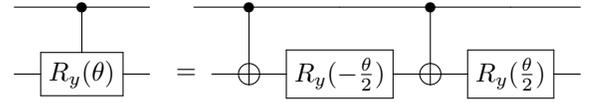
\begin{figure}[h!] $\;$
  \centering
  \Qcircuit @C=1em @R=1.5em {
& \ctrl{1}           &\qw&                                        &\ctrl{1}&\qw                  &\ctrl{1}&\qw&\qw\\
& \gate{R_y(\theta)} &\qw&\push{=\rule{.3em}{0em}}&\targ   &\gate{R_y(-\frac{\theta}{2})}&\targ   &\gate{R_y(\frac{\theta}{2})}&\qw
}
  \caption{Quantum circuit implementing a controlled-$R_y$ gate}
  \label{Ry-QC}
\end{figure}

The training-data oracle in \cite{li2015anexperimental} contains two controlled rotation gates, and when executed on a real quantum machine, each rotation gate needs to be implemented by four elementary gates, as shown in Fig.~\ref{Ry-QC} \cite{schuld2017implementing}. This gives a circuit depth of $9$ for the original training-data oracle.

\begin{figure}[h!]
  \centering
  \includegraphics[width=9cm]{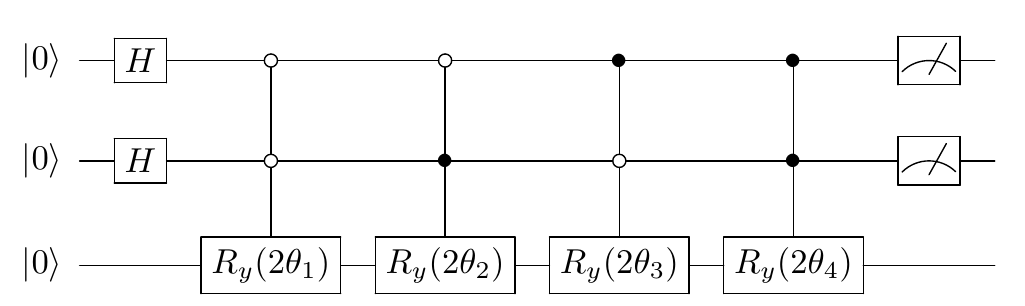}
  \caption{The original training-data oracle when loading four training data}
  \label{original-TDO-4}
\end{figure}

\begin{figure}[h!] 
  \centering
  \includegraphics[width=9cm]{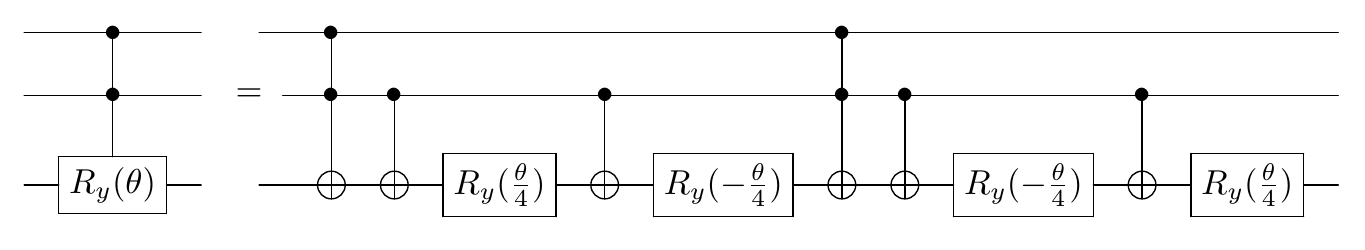}
  \caption{Quantum circuit implementing a $CC-R_y$ gate}
  \label{CCrotation}
\end{figure}

Let us assume that we want to load more training data, for example, four training data. The training-data oracle for four training data is shown in  Fig. \ref{original-TDO-4}, where each controlled controlled rotation gate in this circuit needs to be implemented by $10$ elementary gates, as shown in Fig. \ref{CCrotation} \cite{schuld2017implementing}.

This leads to a circuit depth of $41$ for the circuit in Fig. \ref{original-TDO-4}. As mentioned before, NISQ computers have limited coherence time and a quantum circuit with depth $41$ will fail to be computed on IBMQX2. 
To reduce the circuit depth, we propose a new training-data oracle for generating the kernel matrix $K$, at the expense of using more qubits when loading more than two training data. We would like to design a new training-data oracle to prepare a state $|\phi\rangle$ as dictated by Equation \ref{vector_3}.

\begingroup
\fontsize{8pt}{8pt}\selectfont
\begin{align}
   |\phi\rangle&=\sum_{i=1}^{M}|i-1\rangle\otimes|x_i\rangle \notag
   \\&=|0\rangle\otimes|x_1\rangle+|1\rangle\otimes|x_2\rangle+\cdots+|M-1\rangle\otimes|x_M\rangle\notag
   \\&=\begin{pmatrix}
(\vec{x_1})_1\\ 
(\vec{x_1})_2\\ 
(\vec{x_2})_1\\ 
\vdots\\ 
(\vec{x_M})_1\\ 
(\vec{x_M})_2
\end{pmatrix}\label{vector_3}
\end{align}
\endgroup


\begin{figure}[h!] $\;$
  \centering
  \Qcircuit @C=1em @R=1.5em {
&\lstick{\ket{0}} & \gate{R_y(2\theta_1)} &\meter & \qw& \\
&\lstick{\ket{0}} & \gate{R_y(2\theta_2)} &\meter & \qw& \\
&\lstick{\vdots } &        \vdots          & \rstick{\vdots }  & \\
&\lstick{\ket{0}} & \gate{R_y(2\theta_n)} &\meter & \qw& 
}
  \caption{New training-data oracle of generating $K$}
  \label{optimized-K-QC}
\end{figure}
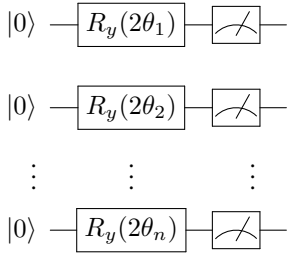

The new training-data oracle is shown in Fig.~\ref{optimized-K-QC}. After running the quantum circuit, we can measure the respective state of each qubit to obtain $M$ two-dimensional results $\begin{pmatrix}
(\vec{x_i})_1\\ 
(\vec{x_i})_2
\end{pmatrix}$,$\ i=1, \ldots, M$, and calculate the tensor product of all two-dimensional results $\begin{pmatrix}
(\vec{x_i})_1\\ 
(\vec{x_i})_2
\end{pmatrix}$ to obtain the state $|\phi\rangle$. 
    
Knowing $|\phi\rangle$, the matrix $\hat{K}$ can be determined by the partial trace of $\frac{1}{M}|\phi\rangle \langle \phi |$:
   
   \begingroup
\fontsize{8pt}{8pt}\selectfont
\begin{equation}
   \hat{K}=K/tr(K)=tr_B(\frac{1}{M}|\phi\rangle \langle \phi |)
\end{equation}
\endgroup

The comparison between the two training-data oracles with regards to the circuit depth and the number of qubits in the circuit is shown in Table \ref{compare-TDO}. As we can see, there is a trade-off between the circuit depth and the number of qubits in the training-data oracles. Since our QSVM system is based on small-scale training data, we choose the new training-data oracle in our implementation.

\begin{table}[h!]
\centering
\begin{tabular}{c|c|c|c|c}
\hline
\multirow{2}{*}{\textbf{\tabincell{c}{number of \\ training data}}} & \multicolumn{2}{c|}{\textbf{\tabincell{c}{circuit depth of the\\training-data oracle}} }                      & \multicolumn{2}{c}{\textbf{\tabincell{c}{number of qubits in \\the training-data oracle}}}    \\ \cline{2-5} 
                                         & original& new&original &new\\ \hline
2                                        & $9$                             & $1$                        & $2$                             & $2$                        \\ \hline
4                                       & $41$                            & $1$                        & $3$                             & $4$                        \\ \hline
\vdots                                   & $\vdots$                        & $\vdots$                   & $\vdots$                        &$\vdots$                    \\ \hline
M                                        & $3M^2-2M+1$                     & $1$                        & $log_2M+1$                              & $M$                        \\ \hline
\end{tabular}
\caption{The comparison of two training-data oracles}
\label{compare-TDO}
\end{table}

The matrix $\hat{K}$ obtained by the new-training data oracle using IBM's quantum computer, IBMQX2, is $\begin{pmatrix}
0.5 &0.240 \\ 
 0.240&0.5
\end{pmatrix}$.
As related in \cite{li2015anexperimental}, $\hat{K}$ is the kernel matrix $K$ divided by $tr(K)$. For our normalized training data, $tr(K)=2$, because the two main diagonal values of the kernel matrix, $K_{jj}$ for $j=1,2$, can be calculated by $K_{jj}=\vec{x_j}\cdot \vec{x_j}=1$ and they sum up to $2$. Thus, the kernel matrix $K$ is equal to $K=2\times \begin{pmatrix}
0.5 &0.240 \\ 
 0.240&0.5
\end{pmatrix}=\begin{pmatrix}
1 &0.48 \\ 
 0.48&1
\end{pmatrix}$. In our implementation, the user-defined parameter $\gamma$ is set to be a relatively big value to shift the weight on the training error to avoid misclassification. By setting $\gamma = 2^3$, as an example, we can then calculate the matrix $F$ by:

\begingroup
\fontsize{8pt}{8pt}\selectfont
\begin{equation}
    F=K+\gamma^{-1}I=\begin{pmatrix}
1+\gamma^{-1} &0.48 \\ 
 0.48&0.5 +\gamma^{-1} 
\end{pmatrix}\approx\begin{pmatrix}
1 &0.5 \\ 
 0.5&1 
\end{pmatrix}
\end{equation}
\endgroup

\subsection{Optimized HHL quantum circuit for QSVM}
To implement QSVM with a short-depth quantum circuit, we modify the optimized HHL circuit provided in reference \cite{cai2013experimental} and propose a result readout method to make the circuit suitable for QSVM classification problems. 

\begin{figure}[h!] $\;$
  \centering
  \includegraphics[width=9cm]{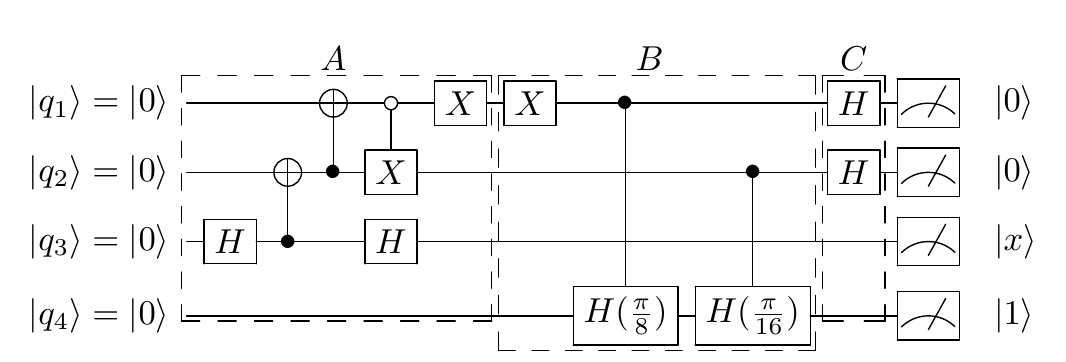}
  \caption{Optimized HHL quantum circuit for QSVM classification. It comprises three parts: part A, phase estimation; Part B, controlled rotation; and Part C, inverse phase estimation. The two $X$ gates can be cancelled out.}
  \label{optimized-HHL-QSVM}
\end{figure}

Reference \cite{cai2013experimental} introduces an optimized quantum circuit of implementing the HHL algorithm, which can reduce the circuit depth from $18$ to $7$. However, the quantum circuit is only designed for the cases where the matrix $\begin{pmatrix}
1.5 & 0.5\\ 
 0.5&1.5 
\end{pmatrix}$ with eigenvalues $1$ and $2$, is used as the input of the HHL algorithm. To make the circuit suitable for our implementation, i.e., where matrix $F=\begin{pmatrix}
1 & 0.5\\ 
 0.5&1 
\end{pmatrix}$ with eigenvalues $0.5$ and $1.5$,  is used as the input of the HHL algorithm, we need to change the original $X$ gate into a controlled-$X$ gate and the original two SWAP gates into two $X$ gates. The modified quantum circuit is shown in Fig. \ref{optimized-HHL-QSVM}.

After the Hadamard gate and the two controlled-NOT gates, the first three qubits become a 3-qubit Greenberger-Horne-Zeilinger state (GHZ) state, $\frac{1}{\sqrt{2}}(|000\rangle+|111\rangle)$. The original design has a $X$ gate on the second qubit, changing the GHZ state into $\frac{1}{\sqrt{2}}(|010\rangle+|101\rangle)$, where the first two qubits encode the eigenvalues of the matrix $\begin{pmatrix}
1.5 & 0.5\\ 
 0.5&1.5 
\end{pmatrix}$, $|\lambda_1\rangle = |01\rangle$ and $|\lambda_2\rangle = |10\rangle$. If we use the control-X gate (the $X$ gate will act on the target qubit when the control qubit is in the state $|0\rangle$) instead, the GHZ state will be changed into $\frac{1}{\sqrt{2}}(|010\rangle+|111\rangle)$, where the first two qubits encode the eigenvalues of the matrix $\begin{pmatrix}
1 & 0.5\\ 
 0.5&1 
\end{pmatrix}$, $|01\rangle$ and $|11\rangle$.  The Control-X gate followed by a $X$ gate and a Hadamard gate, will end the phase estimation part A in Fig. \ref{optimized-HHL-QSVM}. The controlled rotation part B in Fig. \ref{optimized-HHL-QSVM} contains a $X$ gate and two $H(\theta)$ rotations. The $X$ gate function is to find the reciprocals $|\frac{1}{\lambda_i}\rangle, i=1,2$ from the eigenvalues $|\lambda_i\rangle, i=1,2$.
The $H(\theta)$ gate is defined as $H(\theta)=\begin{pmatrix}
cos(2\theta) & sin(2\theta)\\ 
sin(2\theta) & -cos(2\theta)
\end{pmatrix}$ \cite{cai2013experimental}. The final part of the circuit, C in Fig. \ref{optimized-HHL-QSVM} is composed of two Hadamard gates for computing the inverse phase estimation.

The quantum circuit in Fig.~\ref{optimized-HHL-QSVM} solves the matrix inversion problem, $\vec{\alpha}=F^{-1}\vec{y}$. However, we also need to calculate the parameters $\alpha_1$ and $\alpha_2$ of the hyperplane in order to solve QSVM classification problems. The parameters $\alpha_1$ and $\alpha_2$ will define the decision boundary to separate the test data points. They can be calculated as:

\begingroup
\fontsize{8pt}{8pt}\selectfont
\begin{equation}
\begin{aligned}
    \alpha_1=|a_1|\\
    \alpha_2=-|a_3|
\end{aligned}
\end{equation}
\endgroup



where the coefficients $a_1$ and $a_3$ are the amplitudes of  $|0001\rangle$ and $|0011\rangle$ quantum states, respectively, obtained after running  Fig.~\ref{optimized-HHL-QSVM}'s quantum circuit. Once the two parameters of the decision boundary are known, together with the third parameter $b=0$ for our non-offset case, we can classify each test data by Equation \ref{Classification-formula}.

\section{Methodology}

Our QSVM implementation adopts the linear kernel, and it is to solve the non-offset QSVM problems, so the expected datasets should be linear separable by a line crossing the origin. The implementation is tested by two datasets, the OCR dataset and Iris dataset. 

\subsection{Datasets}
\subsubsection{OCR dataset}
\begin{figure}[h!]
  \centering
  \includegraphics[width=8.8cm]{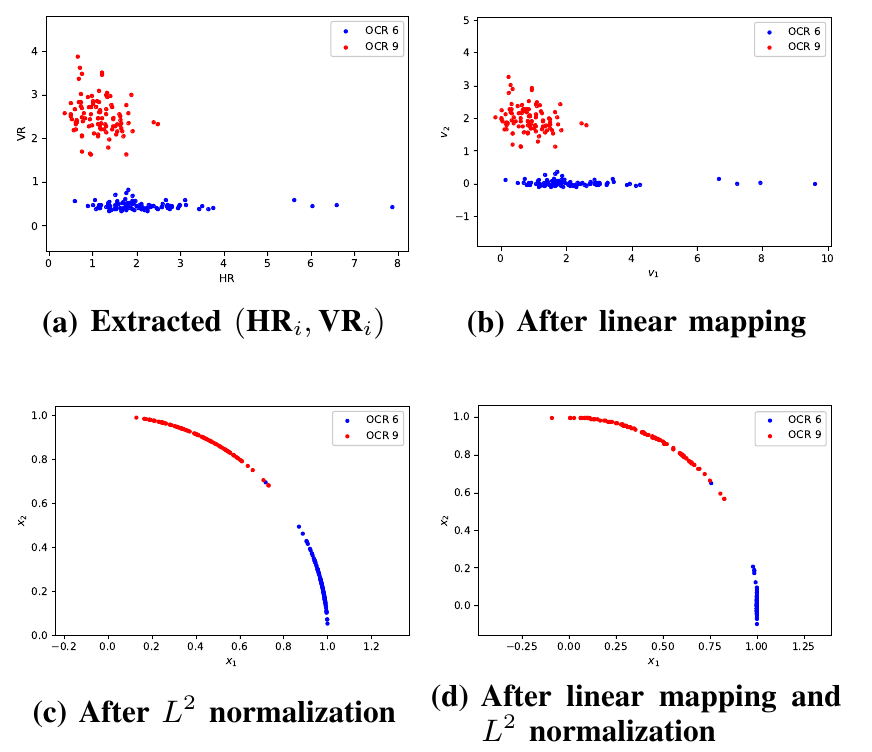}
  \caption{OCR data layout in different steps of preprocessing unit}
  \label{fig:data preprocessing}
\end{figure}

The optical character recognition (OCR) dataset \cite{OCR_dataset} contains the handwritten images of numbers $0$ to $9$, and there are $100$ different images for each number. In our implementation, the printed version of the letters ``6" and ``9" are used as the two training data, and the handwritten ``6" and ``9" images in the OCR dataset are used as test data. Thus, there are $200$ test data in total, half of which belong to the $+1$ class, i.e., handwritten ``6" images, and the other half  belong to the $-1$ class, i.e., handwritten ``9" images. The layout of all test data during different preprocessing steps is shown in Fig.~\ref{fig:data preprocessing}. The blue points refer to data ``6", and the red points refer to data ``9". Fig.~\ref{fig:data preprocessing}a shows the layout of the two-dimensional features, $(\textrm{HR}_i, \textrm{VR}_i)$. Fig.~\ref{fig:data preprocessing}b and Fig.~\ref{fig:data preprocessing}c show the features after linear mapping and normalization, respectively. Fig.~\ref{fig:data preprocessing}d shows the final two-dimensional data $((\vec{x_i})_1, (\vec{x_i})_2)$, which are the data points after linear mapping and normalization.


\subsubsection{Iris dataset}
\begin{figure}[h!]
  \centering
  \includegraphics[width=8.6cm]{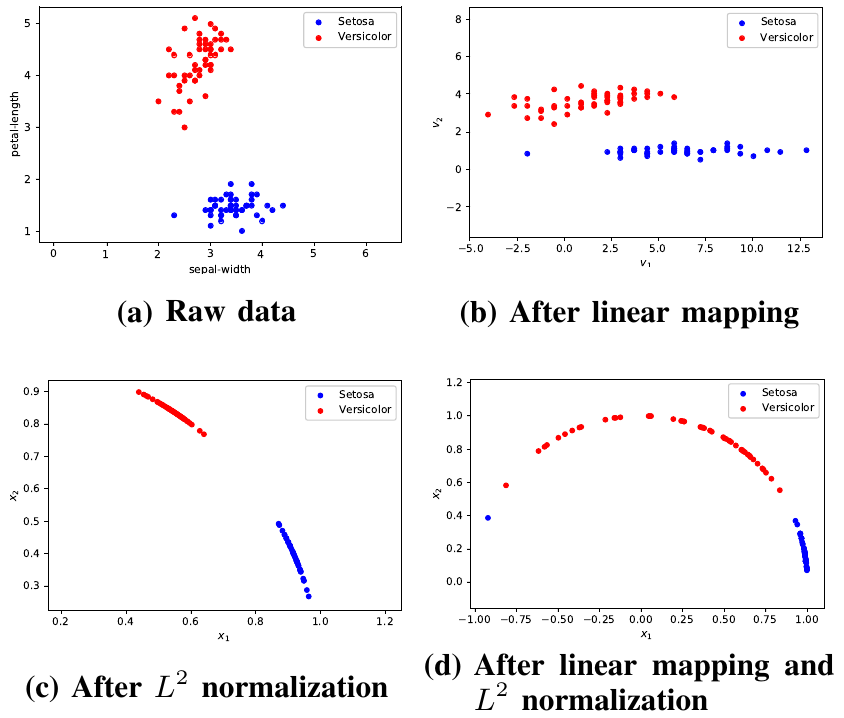}
  \caption{Iris data layout in different steps of preprocessing unit}
  \label{fig:Iris dataset}
\end{figure}

The Iris dataset ~\cite{UCI-Iris} contains $50$ samples from three different species of the iris flower: ``setosa", ``versicolor" and ``virginica". Each data in the dataset has four dimensions, ``sepal.length", ``sepal.width", ``petal.length" and ``petal.width". Our QSVM system is currently a non-offset system and limited to two-dimensional test data that can be linearly separated into two subgroups by a line crossing the origin. Thus, we choose only two classes, ``setosa" and ``versicolor", and two dimensions from the original data: ``sepal.width" as the first dimension of the test data, and ``petal.length" as the second dimension. In total, there are $100$ test data, half of which belong to the $+1$ class, i.e., ``setosa", and the other half of which belong to the $-1$ class, i.e., ``versicolor". The initial layout of these $100$ test data points is shown in Fig.~\ref{fig:Iris dataset}a.

The mean value of all data in $+1$ class is designed to be the first training data, and similarly, the mean value of all data in $-1$ class is designed to be the second training data. Both the training data and test data should  be linearly mapped first, using the new coefficients calculated by Equations~\ref{fangcheng_1} to~\ref{fangcheng_4}, then normalized and converted into rotation angles. The test data after linear mapping is shown in Fig. \ref{fig:Iris dataset}b, and the normalized test data without and with the linear mapping is shown in Fig.~\ref{fig:Iris dataset}c and~\ref{fig:Iris dataset}d.


\subsection{Executing platforms and hardware}
In our implementation, two quantum platforms, pyQuil and Qiskit, are used to execute the quantum circuits. All simulations are run on pyQuil, while all real quantum computer executions are run on Qiskit.
\subsubsection{pyQuil and the Wavefunction Simulator}
pyQuil \cite{PyQuil} is a quantum computing software development kit (SDK) provided by a start-up quantum computing company, Rigetti. It is Python-based, and users can test their quantum algorithms by simulating quantum circuits and also executing quantum programs on Rigetti's  actual quantum harwdare.

The Wavefunction Simulator is pyQuil's tool to provide the wave function. A wave function is a mathematical description of the quantum state of a quantum system. It comprises complex-valued amplitudes, and we can derive the probabilities of the possible measurement states from the amplitudes. 
 Different from quantum computers or local simulators, which have to run the quantum programs for many iterations to get the probability distribution of the quantum states, the Wavefunction Simulator enables us to get the theoretical amplitude distribution of the quantum states directly after running the quantum program once. 

In this paper, the simulations of all quantum circuits are implemented by pyQuil's Wavefunction Simulator, because the quantum circuits involve some gates that are not commonly used, and pyQuil allows us to define our own gates with unitary matrices. 

\subsubsection{Qiskit and IBMQX2 quantum computer}

Qiskit \cite{Qiskit} is a Python-based open-source SDK for creating quantum computing programs provided by IBM. It works with the quantum language OpenQASM. Users can design quantum circuits, simulate quantum programs by using their host computers as local simulators, or even get access to IBM Q Experience \cite{IBMQexperience}'s  remote quantum computers through this SDK. Qiskit also provides an error-correction tool, Ignis \cite{Ignisgithub}, to characterize and mitigate part of the noise present in quantum circuits and devices, and that is why we choose IBM's quantum hardware. 

In this paper, all quantum algorithms running on a real quantum computer are implemented by Qiskit, error-correction tool Ignis, and IBM Q Experience's 5-qubit superconducting quantum computer, IBMQX2. Currently, Qiskit does not support user-defined gates, so some uncommon gates, such as the $H(\theta)$ gate in Fig. \ref{optimized-HHL-QSVM},  can not be directly implemented by Qiskit.  To solve this problem, we first compile the gate in pyQuil, and then apply the elementary gates after compilation to Qiskit. Each algorithm is executed for $8192$ iterations, in order for us to obtain the probability distribution of the quantum states. 

\begin{figure}[h!] 
  \centering
  \includegraphics[width=3cm]{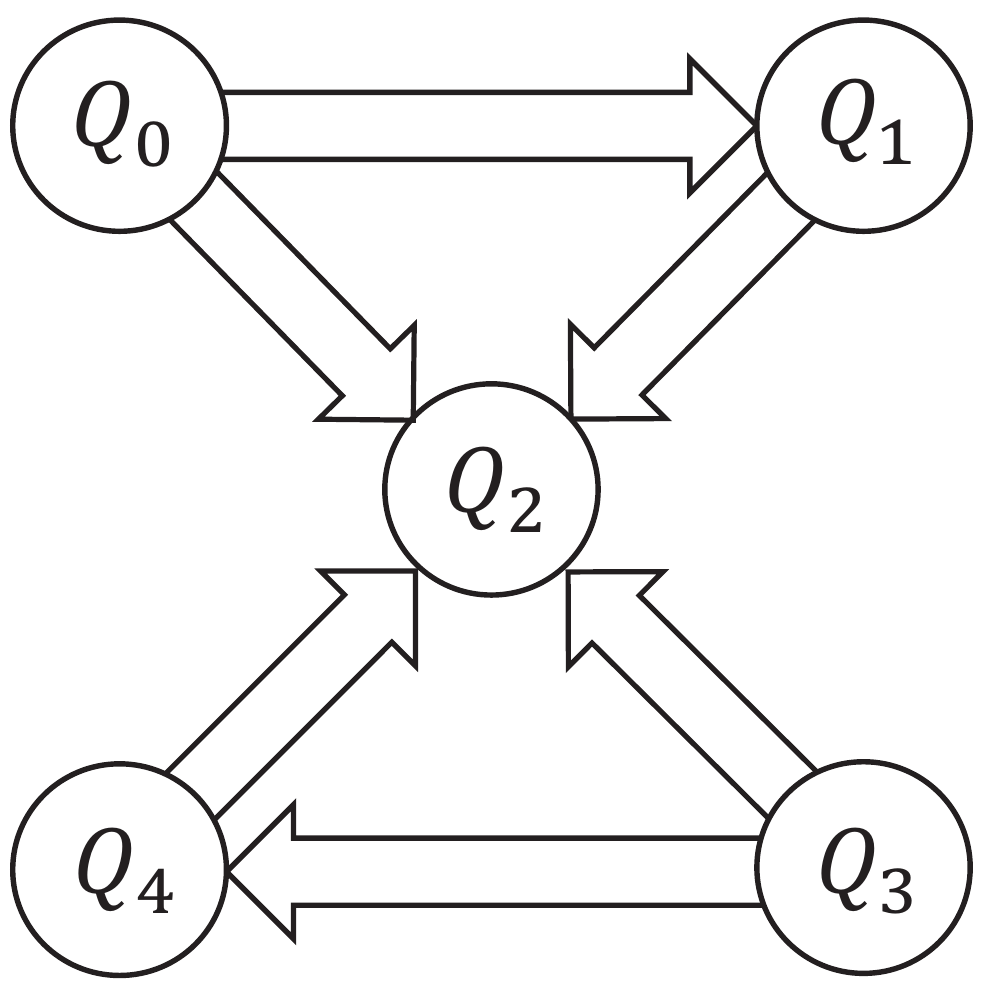}
  \caption{The architecture of IBM Q Experience's 5-qubit quantum computer, IBMQX2, cited from \cite{dueck2018optimization}}
  \label{IBMQX2}
\end{figure}

When executing quantum circuits on IBMQX2, we also need to consider the quantum chip interconnect architecture. The IBMQX2 chip interconnect is shown in Fig. \ref{IBMQX2}. As described in \cite{dueck2018optimization} a CNOT gate can only be implemented between two connected qubits, and a qubit at the tail of an arrow can only be the control qubit, while the qubit at the head of the arrow can only be the target qubit. When running the quantum circuit depicted in Fig. \ref{optimized-HHL-QSVM}'s  on IBMQX2, we will need to consider the qubits interconnections. Let us consider the two CNOT gates needed to prepare the GHZ state. In this case, we can use IBMQX2's $Q_2$ qubit to implement the $|q_1\rangle$ of the algorithm; IBMQX2's $Q_1$ qubit to implement the $|q_2\rangle$; IBMQX2's $Q_0$ qubit to implement the $|q_3\rangle$; and IBMQX2's $Q_3$ qubit to implement the $|q_4\rangle$ of the algorithm. 

\subsection{Metric for measuring the similarity between two probability distributions}
Our implementation uses relatively small-scale training data, so the classification accuracy itself sometimes is not enough to evaluate the performance of a quantum circuit. Jensen-Shannon (JS) divergence \cite{endres2003new} is a metric to calculate the difference between two probability distributions, and it can also be used to measure the  distinguishability between two quantum states \cite{majtey2005jensen}. To better evaluate the performance of a quantum circuit, especially when the circuit runs on a noisy quantum computer, we introduce JS divergence as another metric to calculate the difference between the probability distribution obtained from the IBMQX2 and the theoretical probability distribution from the local simulator. JS divergence is chosen here as it can overcome the Kullback-Leibler (KL) divergence's drawback of asymmetry. 

Let us assume that we have two probability distributions $P_1$ and $P_2$, then the JS divergence between them can be calculated by:

\begingroup
\fontsize{8pt}{8pt}\selectfont
\begin{equation}
D_{JS}(P_1||P_2)=\frac{1}{2}D_{KL}(P_1||\frac{P_1+P_2}{2})+\frac{1}{2}D_{KL}(P_2||\frac{P_1+P_2}{2})
\end{equation}
\endgroup

where $D_{KL}$ stands for the KL divergence, defined as:

\begingroup
\fontsize{8pt}{8pt}\selectfont
\begin{equation}
D_{KL}(P_1||P_2)=-\sum_iln\frac{P_1(i)}{P_2(i)}
\end{equation}
\endgroup

The JS divergence, fulfilling $D_{JS}(P_1||P_2)=D_{JS}(P_2||P_1)$,  is a symmetric metric for measuring the similarity between two probability distributions. The JS divergence between any two probability distributions is always in the range of $[0, 1]$. The closer the JS divergence is to $0$, the more similar the two probability distributions are.

\subsection{Optimized baseline}
In reference \cite{li2015anexperimental}, the training and testing process provided is too time-consuming to be implemented on a quantum computer on the cloud, especially when there are a lot of test data. For every new-coming test data, the QSVM classification circuit (circuit depth is $22$) needs to be run once to get the classification result. Assuming we have $m$ test data points, both the training part of the circuit (the matrix inversion), and the testing part of the circuit (the training data oracle and the $U_{x_0}$) of the circuit need to be run for $m$ times. This causes a lot of redundancy, and the total circuit depth for $m$ test data $x_0$ is $(20+2)\times m=22m$.

\begin{figure}[h!]$\;$
  \centering
  \includegraphics[width=8cm]{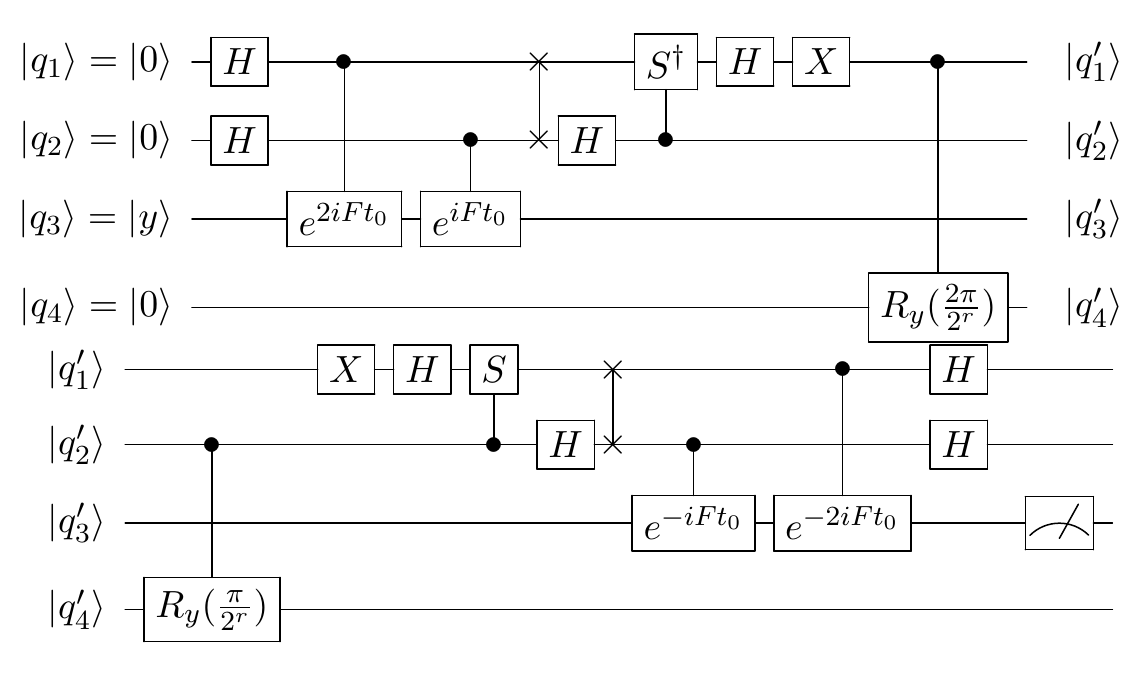}
  \caption{The quantum circuit of QSVM classification, as part of the optimized baseline}
  \label{QSVM_circuit_classical_test}
\end{figure}

To remove the redundancy and in turn, the circuit depth, we propose a new process of testing $optimized baseline$, that together with the preprocessing unit and the kernel matrix generation, it is used as a baseline for the comparison with our QSVM implementation. For this optimized baseline, a new quantum circuit without the testing part is proposed, and it is shown in Fig.~\ref{QSVM_circuit_classical_test}. Note that the measurement is changed from the fourth qubit $|q_4\rangle$ to the third qubit $|q_3\rangle$. Instead of running the whole QSVM classification circuit $m$ times, we only need to run the quantum circuit in Fig.~\ref{QSVM_circuit_classical_test} once, and use the classical result readout method to classify all test data into two classes. 

The total circuit depth of this optimized baseline is $18$, which is a $99\%$ reduction on the total circuit depth for $m=100$. This optimized baseline uses classical result readout methods to replace part of the quantum circuit and accelerate the execution process of the QSVM system. In the following, we will describe the classical result readout method.



After running the quantum circuit depicted in Fig. \ref{QSVM_circuit_classical_test} the measured result of all four qubits can be represented by the equation:

\begingroup
\fontsize{8pt}{8pt}\selectfont
\begin{equation}
    |\psi\rangle=|q_1q_2q_3q_4\rangle=a_0|0000\rangle+a_1|0001\rangle+\cdots+a_{15}|1111\rangle
    \label{16_Q_states}
\end{equation}
\endgroup

where $a_0, a_1, \ldots, a_{15}$ are the amplitudes of the quantum basis states $|0000\rangle$ to $|1111\rangle$.

If we define the third qubit (the measured qubit) to be $|q_3 \rangle=\alpha_1 |0\rangle+\alpha_2 | 1\rangle$, then $\alpha_1$ and $\alpha_2$ can be calculated by:

\begingroup
\fontsize{8pt}{8pt}\selectfont
\begin{equation}
\begin{aligned}
    \alpha_1=a_0+a_1+a_4+a_5+a_8+a_9+a_{12}+a_{13}\\
    \alpha_2=a_2+a_3+a_6+a_7+a_{10}+a_{11}+a_{14}+a_{15}
\end{aligned}
\end{equation}
\endgroup

because the sum of all the coefficients of the quantum states that can be represented by $| q_1 q_2 0q_4 \rangle$ is equal to $(a_0+a_1+a_4+a_5+a_8+a_9+a_{12}+a_{13})$, and similarly, the sum of all coefficients that can be represented by $| q_1 q_2 1q_4 \rangle$ is equal to $(a_2+a_3+a_6+a_7+a_{10}+a_{11}+a_{14}+a_{15})$. Knowing $\alpha_1$ and $\alpha_2$, we can use Equation \ref{Classification-formula} to classify each test data.

The optimized baseline can greatly reduce the execution time on the simulator while ensuring the accuracy, as shown in Table \ref{tab: new testing process}. 

\begin{table}[h!]
\includegraphics[width=9cm]{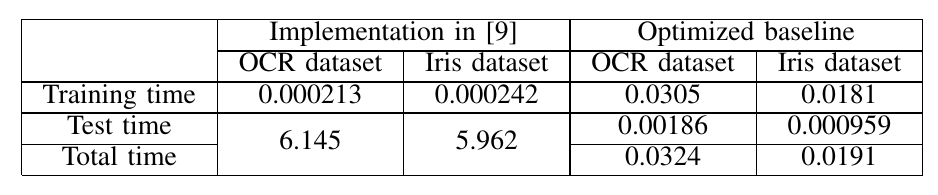}
\caption{The comparison of QSVM execution time on both datasets, using \cite{li2015anexperimental}'s implementation and the optimized baseline}
\label{tab: new testing process}
\end{table}

The readout of the third qubit's coefficients in the optimized circuit makes the training time longer, but the test time and total time are greatly reduced from $6.145$ seconds to $0.0324$ seconds for the OCR dataset, and from $5.962$ seconds to $0.0191$ seconds for the Iris dataset.
\section{Results and discussion}

\subsection{Results on the Pyquil's Wavefunction Simulator}
\subsubsection{Simulated classification results of the QSVM system in the prior art}

In this work, the QSVM system provided by reference \cite{li2015anexperimental} is only re-implemented on the simulator, but not on the quantum computer IBMQX2, because for every new test data, the QSVM classification circuit needs to be run again and the whole process is too time-consuming. What is more, as shown in the following sections,  the optimized baseline's circuit in Fig. \ref{QSVM_circuit_classical_test} is already too long to converge to a successful result on a real superconducting quantum computer, so there is no need for running the more complex quantum circuit in \cite{li2015anexperimental} on IBMQX2.

\begin{figure}[h!]
    \centering
    \begin{subfigure}[b]{0.23\textwidth}
        \centering
        \includegraphics[width=\textwidth]{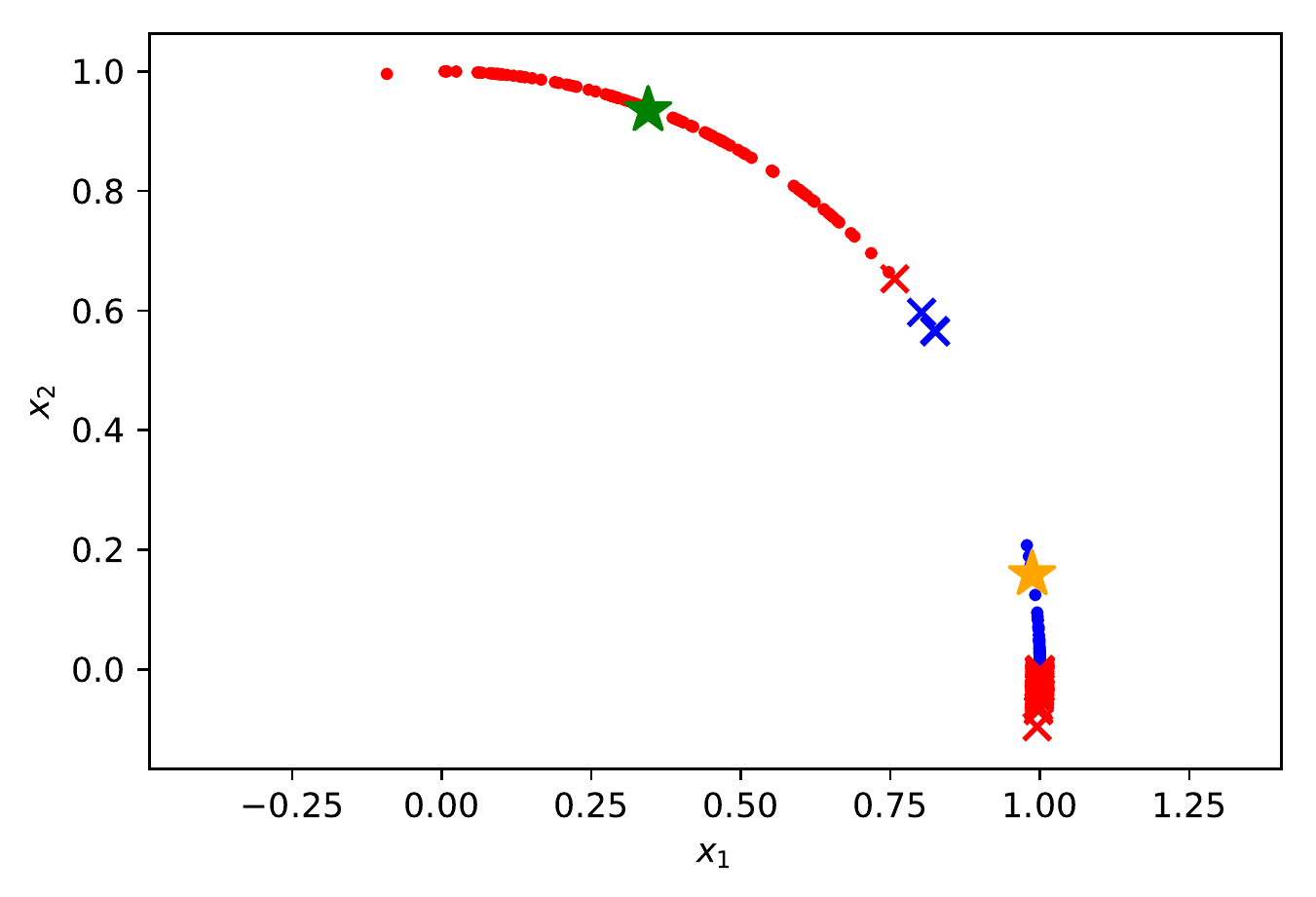}
        \caption[Network2]%
        {{\small OCR dataset}}    
        \label{OCR_result}
    \end{subfigure}
    \begin{subfigure}[b]{0.23\textwidth}   
        \centering 
        \includegraphics[width=\textwidth]{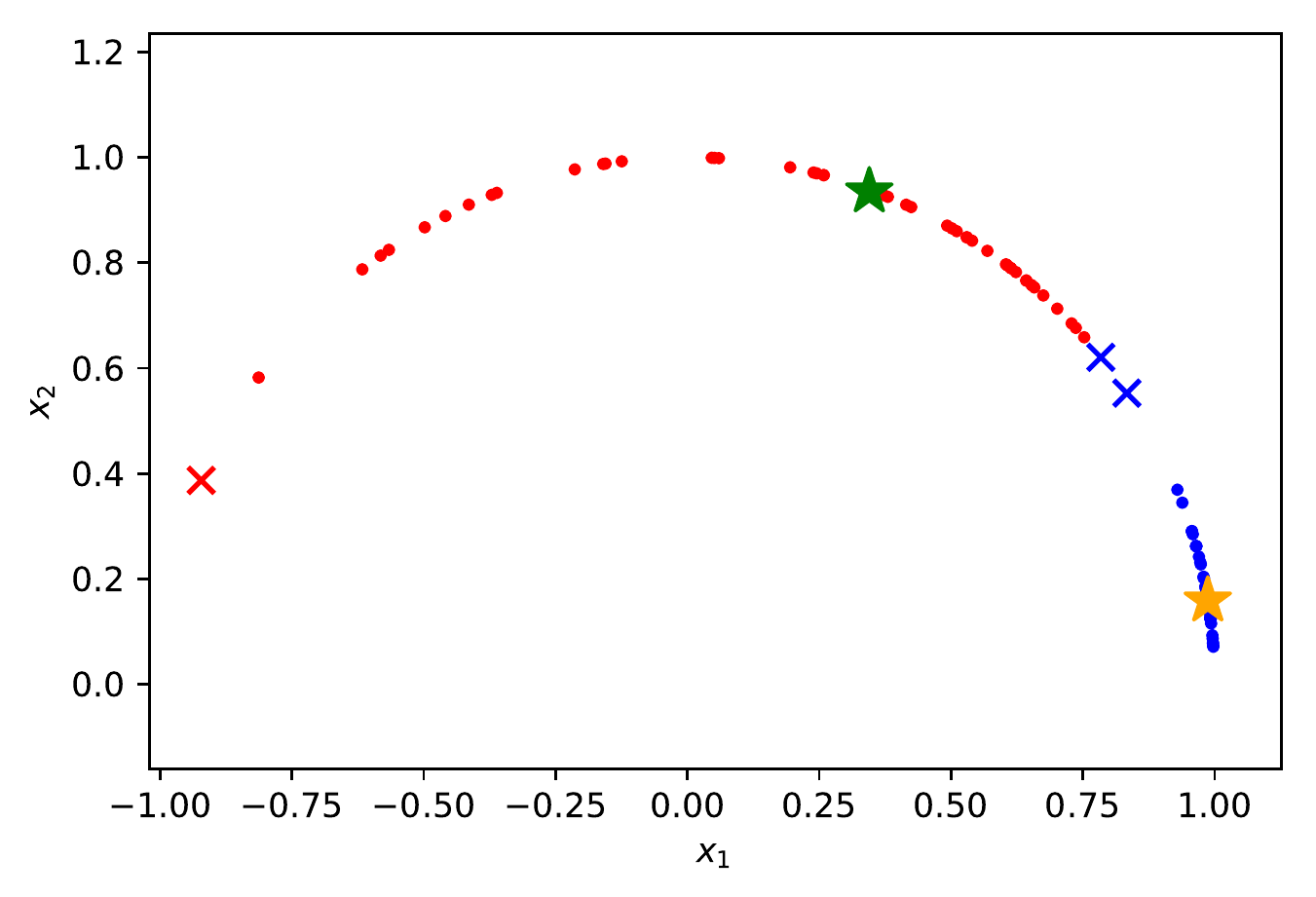}
        \caption[]%
        {{\small}Iris dataset}    
        \label{Iris-result}
    \end{subfigure}
    \caption[Classification results of the QSVM system in reference \cite{li2015anexperimental}]
    {\small Classification results of the QSVM system in reference \cite{li2015anexperimental}} 
\end{figure}

The classification result of 200 OCR test data running on the Wavefunction Simulator of pyQuil is shown in Fig.~\ref{OCR_result}. The blue points stand for the data classified into ``6", while the red points stand for the data classified into ``9". The green and orange stars are the two training data. The test data that are correctly classified are denoted by the dot points ``$\cdot$", while the data that are wrongly classified is denoted by the cross ``$\times$". The classification accuracy is calculated by the percentage of test data classified correctly, which in this case is $74.5\%$. The test data in the fourth quadrant are wrongly classified because, in the preprocessing unit, the $arccot()$ function to calculate rotation angles only works for the data in the first or second quadrants. 

The classification result of the Iris dataset on pyQuil's Wavefunction Simulator is shown in Fig. \ref{Iris-result}. The blue points stand for the test data classified into the $+1$ class, "setosa", and the red points stand for the test data classified into the $-1$ class, "versicolor". The data that is correctly classified is denoted by the dot points "$\cdot$", while the data that is wrongly classified is denoted by the cross "$\times$". The green and orange stars are the two training data. The classification accuracy is $97\%$.

\subsubsection{Simulated classification results of the optimized baseline}

\begin{figure}[h!]
    \centering
    \begin{subfigure}[b]{0.23\textwidth}
        \centering
        \includegraphics[width=\textwidth]{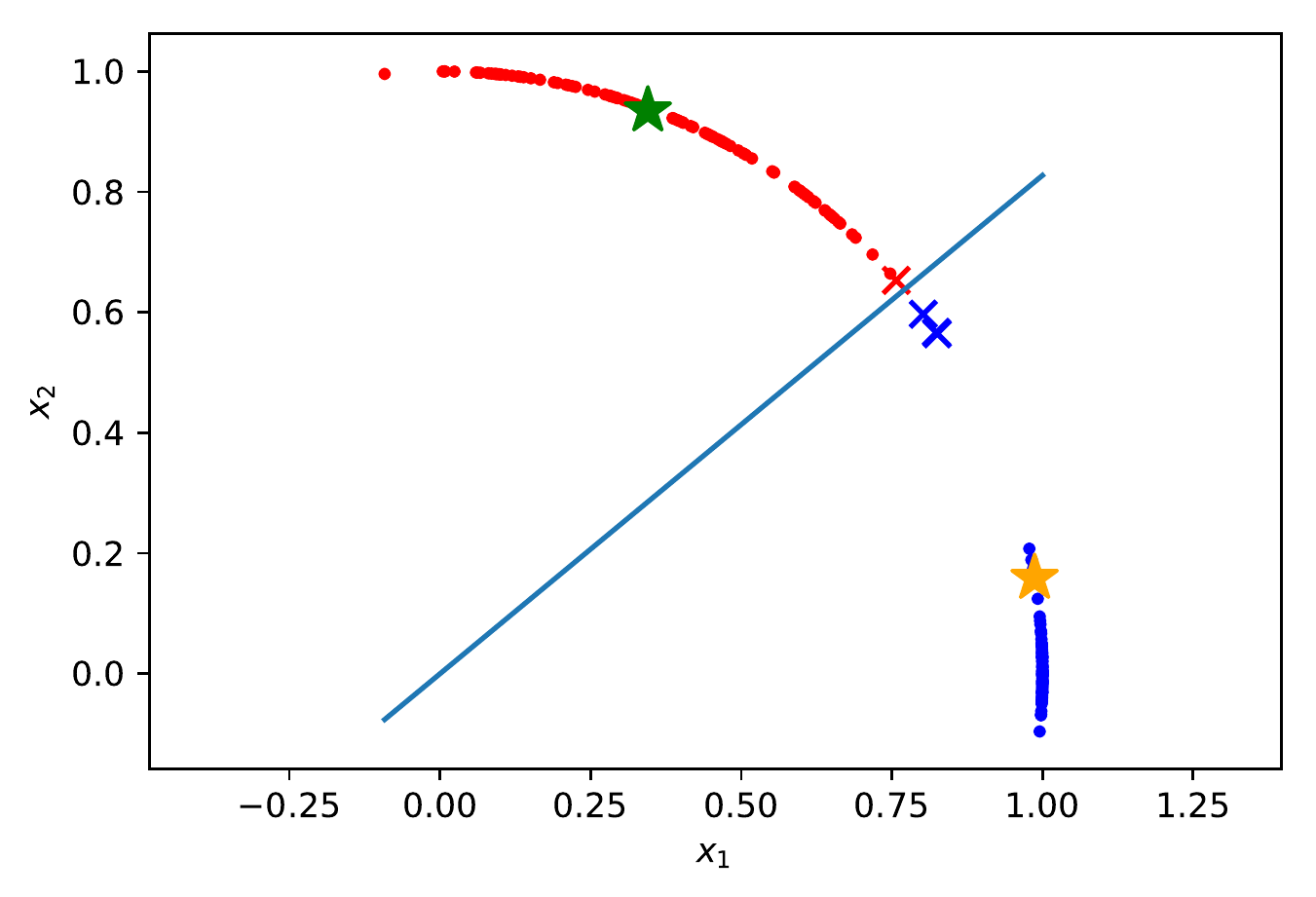}
        \caption[Network2]%
        {{\small OCR dataset}}    
        \label{OCR_result_new_testing}
    \end{subfigure}
    \begin{subfigure}[b]{0.23\textwidth}  
        \centering 
        \includegraphics[width=\textwidth]{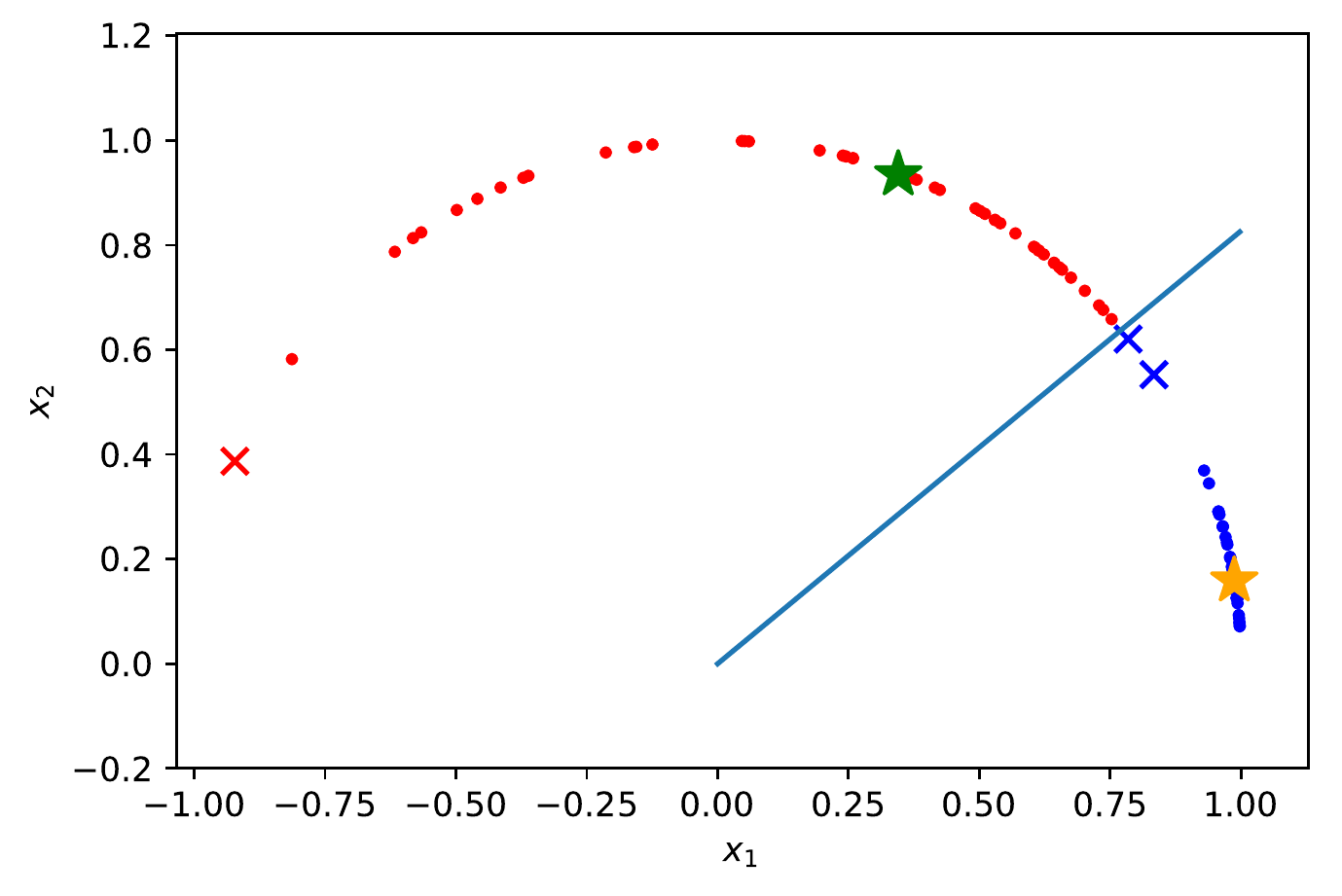}
        \caption[]%
        {{\small Iris dataset}}    
        \label{Iris_result_new_testing}
    \end{subfigure}
    \caption[ Classification results of the optimized baseline]
    {\small Classification results of the optimized baseline} 
\label{fig:new testing}
\end{figure}

The classification results using the optimized baseline and the quantum circuit in Fig. \ref{QSVM_circuit_classical_test} can be found in Fig. \ref{fig:new testing}, where the blue lines denote the decision boundary. The results for the OCR dataset are shown in Fig. \ref{OCR_result_new_testing}. Compared to the results of reference \cite{li2015anexperimental} QSVM system in Fig. \ref{OCR_result} with accuracy $74.5\%$, the optimized baseline has an accuracy $98\%$, because of the new combined set of equations defined in Equation \ref{combineequation} are  suitable for data points in all quadrants.

The results for the Iris dataset are shown in Fig. \ref{Iris_result_new_testing}. Compared to the results shown in ref. \cite{li2015anexperimental} (Fig. \ref{Iris-result}), the accuracy of the optimized baseline remains the same, $97\%$.

\subsubsection{Simulated classification results of the optimized HHL quantum circuit for QSVM classification}
\begin{figure}[h!]
    \centering
    \begin{subfigure}[b]{0.23\textwidth}
        \centering
        \includegraphics[width=\textwidth]{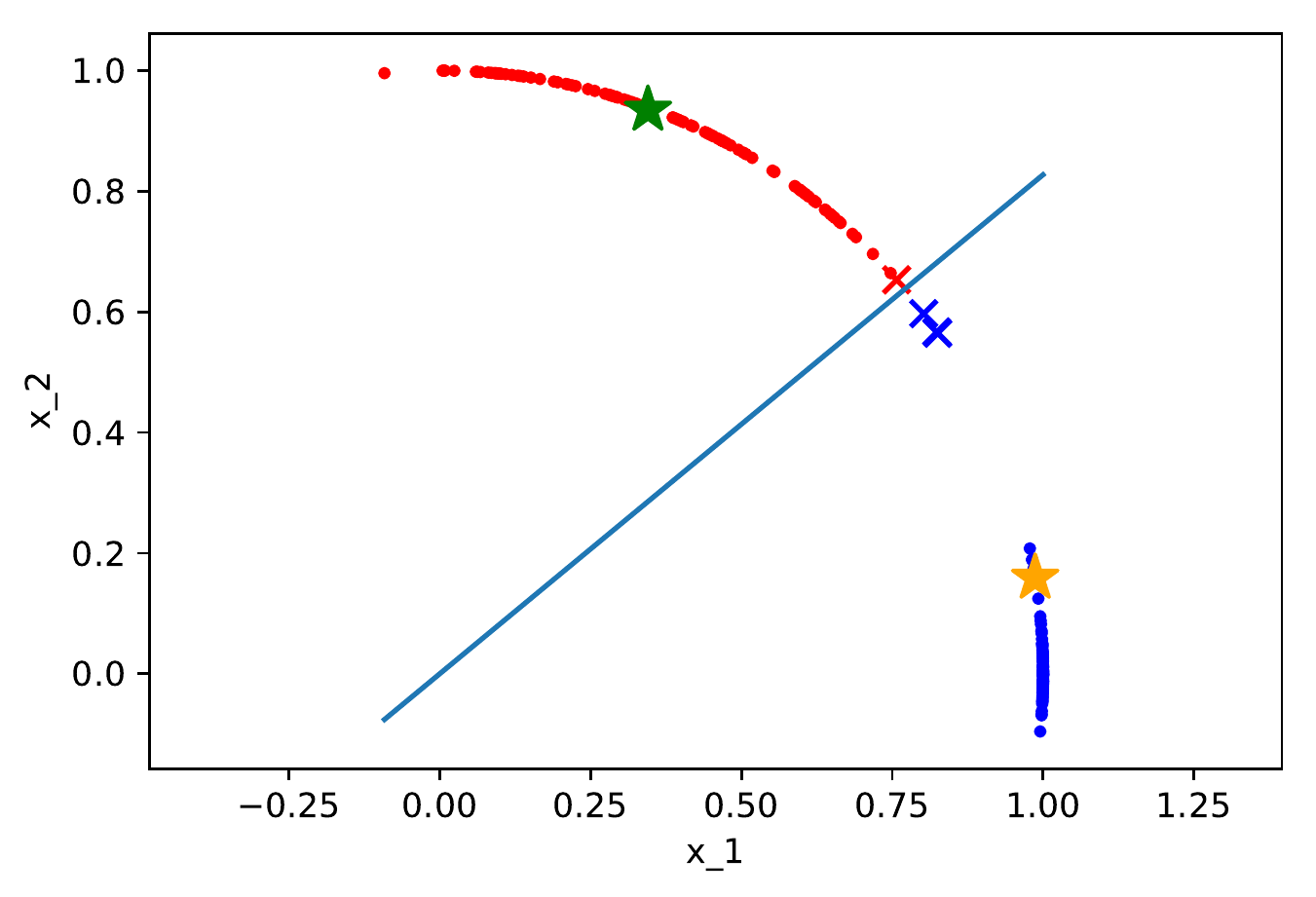}
        \caption[Network2]%
        {{\small OCR dataset}}    
        \label{OCR_result_optimized_simulator}
    \end{subfigure}
    \begin{subfigure}[b]{0.23\textwidth}  
        \centering 
        \includegraphics[width=\textwidth]{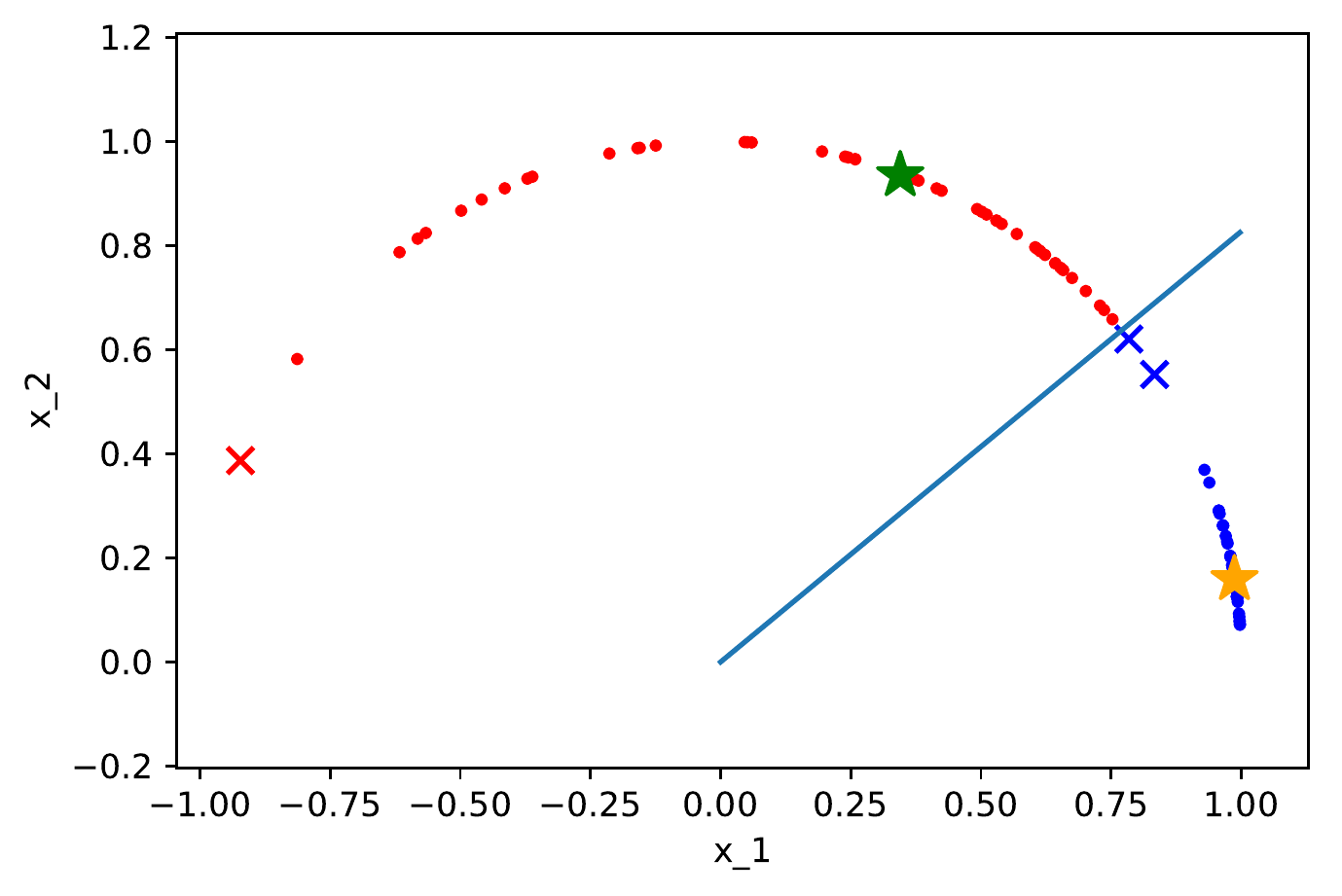}
        \caption[]%
        {{\small Iris dataset}}    
        \label{Iris_result_optimized_simulator}
    \end{subfigure}
    \caption[Classification results of our QSVM implementation]
    {\small Classification results of our QSVM implementation} 
    \label{fig: optimized HHL results}
\end{figure}

The classification results using the optimized HHL quantum circuit in Fig. \ref{optimized-HHL-QSVM} are shown in Fig. \ref{fig: optimized HHL results}. Fig. \ref{OCR_result_optimized_simulator} shows the results on the OCR dataset, with $98\%$ accuracy. Fig. \ref{Iris_result_optimized_simulator} shows the results for the Iris dataset with $97\%$ accuracy. 

\subsection{Results on IBMQX2}
\begin{figure}[h!]
    \centering
    \begin{subfigure}[b]{0.23\textwidth}
        \centering
        \includegraphics[width=\textwidth]{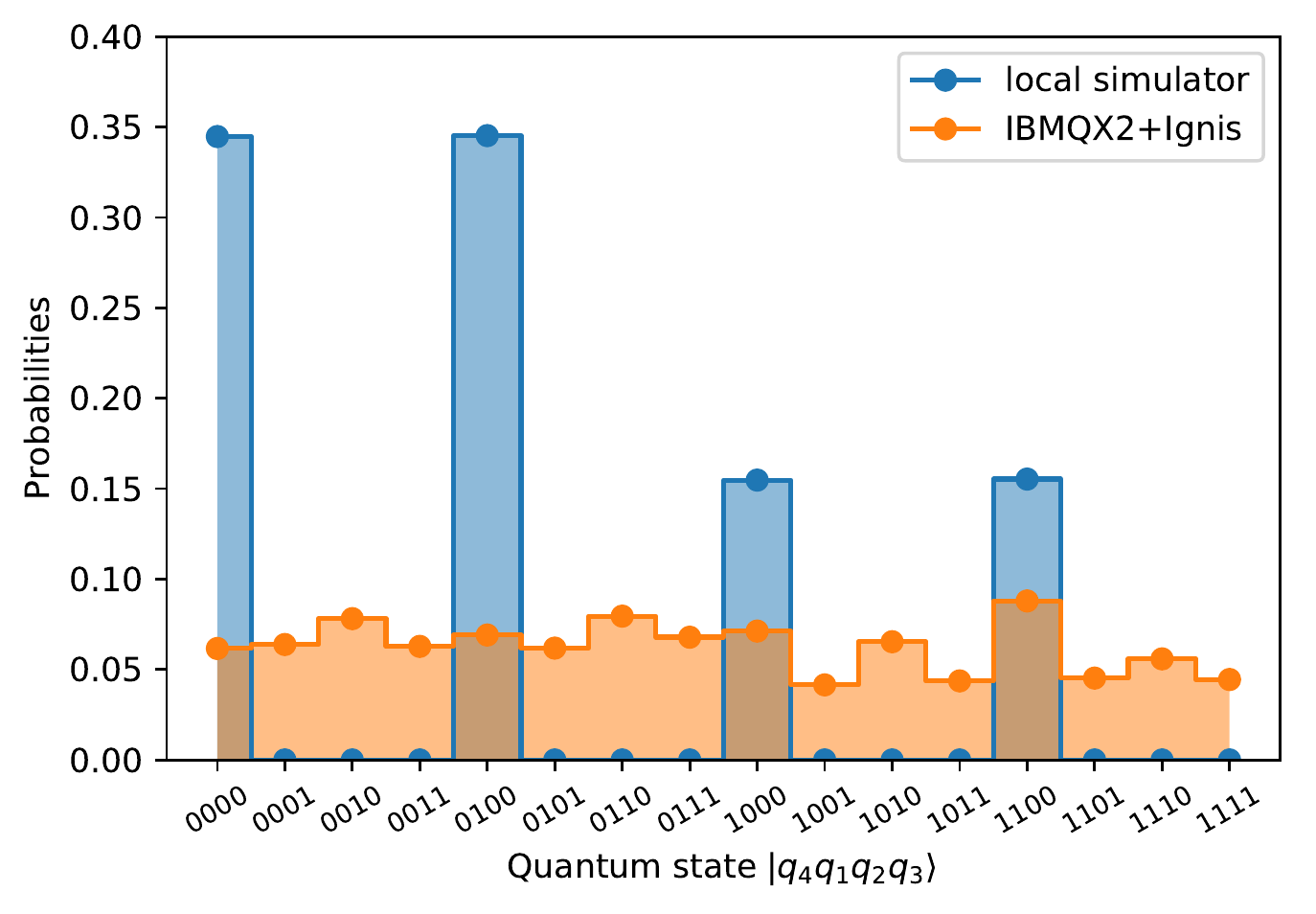}
        \caption[Network2]%
        {{\small Quantum circuit in Fig. \ref{QSVM_circuit_classical_test}}}    
        \label{QSVM_original_QC}
    \end{subfigure}
    \begin{subfigure}[b]{0.23\textwidth}  
        \centering 
        \includegraphics[width=\textwidth]{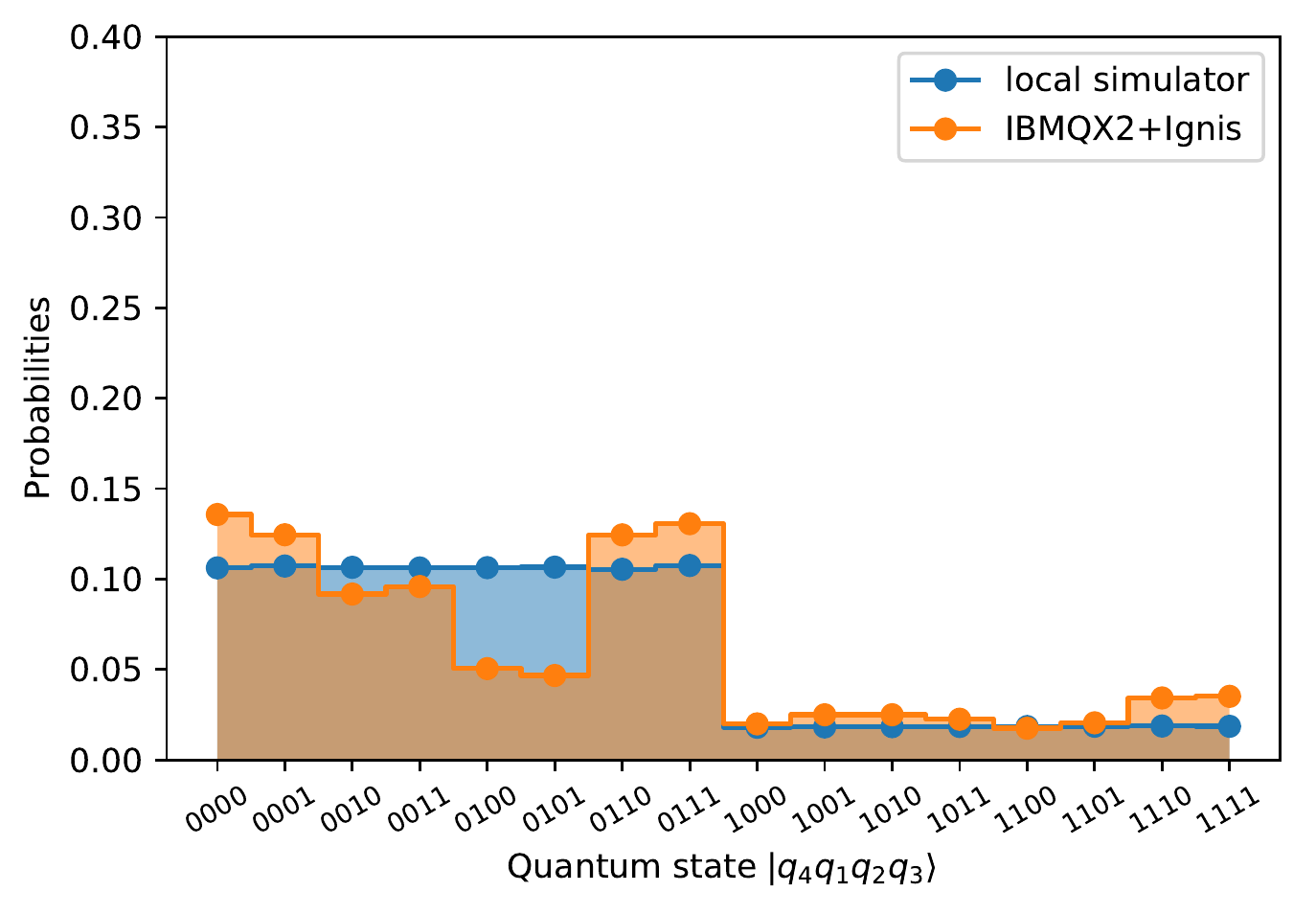}
        \caption[]%
        {{\small Quantum circuit in Fig. \ref{optimized-HHL-QSVM}}}    
        \label{QSVM_optimized_QC}
    \end{subfigure}
    \caption[Probability distributions of the state after quantum circuits running on IBMQX2]
    {\small Probability distributions of the state after running quantum circuits on IBMQX2} 
\end{figure}

Although the optimized baseline method can significantly reduce the execution time of QSVM algorithm, as well as maintain the accuracy, the quantum circuit in Fig. \ref{QSVM_circuit_classical_test} (circuit depth $20$) is still too complicated to run on a current quantum computer. As shown in Fig. \ref{QSVM_original_QC}, the probability distribution of running the quantum circuit in Fig. \ref{QSVM_circuit_classical_test} on IBMQX2 is very different from that on a local simulator, which has no noise or decoherence error.

The optimized HHL quantum circuit for QSVM classification shown in Fig. \ref{optimized-HHL-QSVM} has a circuit depth of $7$. As shown in Fig. \ref{QSVM_optimized_QC}, the probability distribution of running the quantum circuit  of Fig. \ref{QSVM_circuit_classical_test} on IBMQX2 is more similar to the probability distribution obtained on a local simulator, compared to Fig. \ref{QSVM_original_QC}. All the cases are executed $20$ times to decrease the random error before drawing the figures.

We calculate the JS divergence between the probability distributions obtained from IBMQX2 and the local simulator, for both quantum circuits in Fig. \ref{QSVM_circuit_classical_test} and Fig. \ref{optimized-HHL-QSVM}. We observe that the JS divergence of the quantum circuit in Fig. \ref{QSVM_circuit_classical_test} is $0.603$, and that of Fig. \ref{optimized-HHL-QSVM}'s circuit is $0.130$, which means that the probability distribution of running the quantum circuit in Fig. \ref{optimized-HHL-QSVM} is closer to its ideal distribution as shown from the simulator.

\begin{figure}[h!]
    \centering
    \begin{subfigure}[b]{0.23\textwidth}
        \centering
        \includegraphics[width=\textwidth]{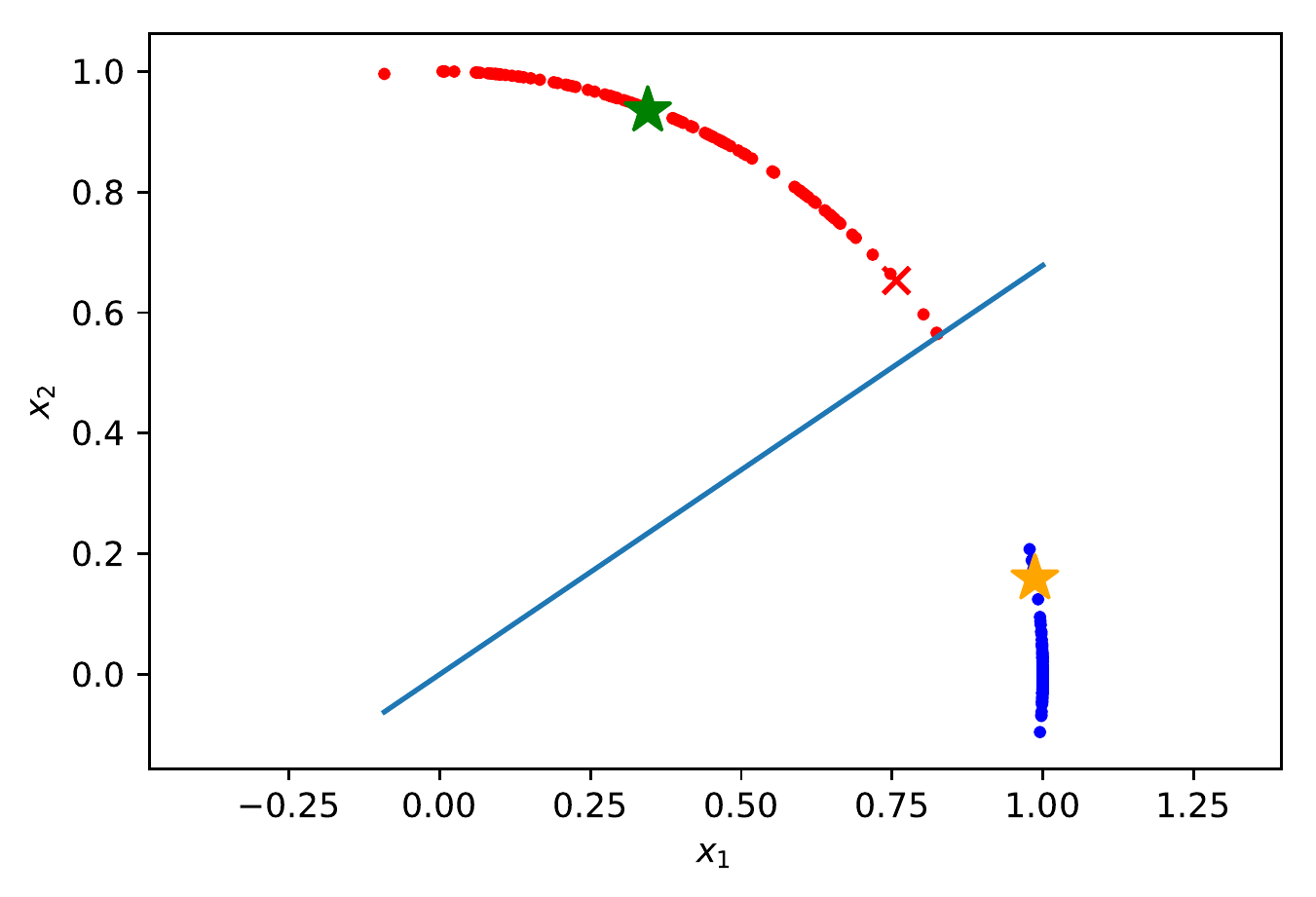}
        \caption[Network2]%
        {{\small OCR dataset}}    
        \label{OCR_result_optimized_QX2}
    \end{subfigure}
    \begin{subfigure}[b]{0.23\textwidth}  
        \centering 
        \includegraphics[width=\textwidth]{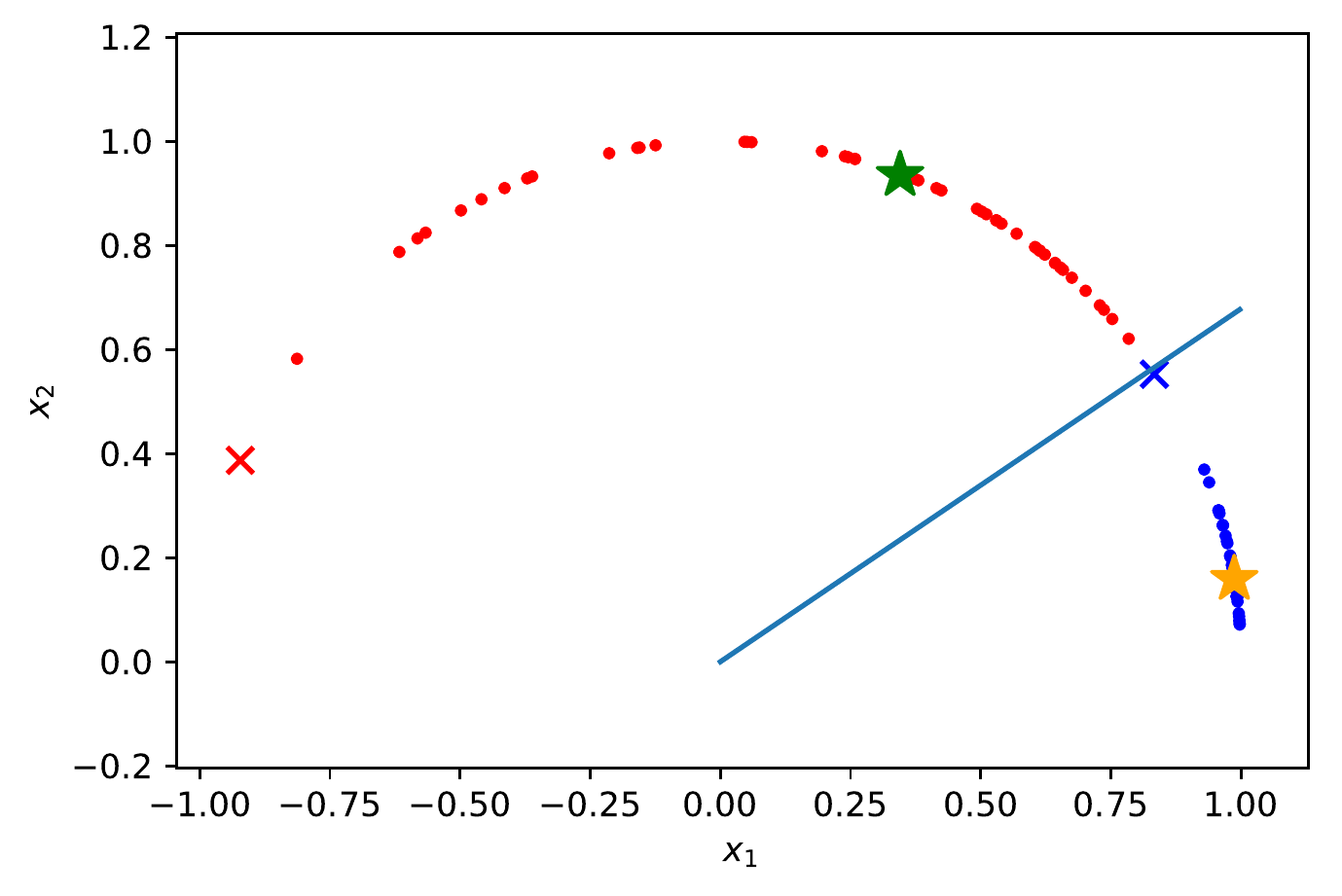}
        \caption[]%
        {{\small Iris dataset}}    
        \label{Iris_result_optimized_QX2}
    \end{subfigure}
    \caption[Classification results using the optimized HHL quantum circuit for QSVM, running on IBMQX2]
    {\small Classification results using the optimized HHL quantum circuit for QSVM, running on IBMQX2} 
    \label{fig: optimized HHL results QX2}
\end{figure}

The classification results on IBMQX2 quantum computer using the optimized HHL quantum circuit in Fig. \ref{optimized-HHL-QSVM} are shown in Fig. \ref{fig: optimized HHL results QX2}. Fig. \ref{OCR_result_optimized_QX2} shows the results on the OCR dataset, with $99.5\%$ accuracy. Fig. \ref{Iris_result_optimized_QX2} shows the results on the Iris dataset with $98\%$ accuracy. 

\section{Conclusions}
In this work, we implemented a new QSVM algorithm that leads to better classification results for the OCR dataset and Iris dataset on both the simulator and the real quantum computer IBMQX2. The implementation comprises three parts, preprocessing unit, the kernel matrix generation and the optimized HHL quantum circuit for QSVM classification. Specifically, we derive an optimized preprocessing unit for a quantum SVM that allows classifying any datasets. We further provide a result readout method of the kernel matrix generation circuit to avoid quantum tomography that in turn, reduces the quantum circuit depth. We also derive a quantum SVM system based on an optimized HHL quantum circuit with reduced circuit depth.
After implementing quantum circuits with shorter depth, we observe more accurate results on actual quantum computers, by reducing the JS divergence between probability distributions of an actual quantum computer and the ideal simulator by $78.4\%$. This points towards the great potential of using NISQ computers to compute machine learning algorithms like SVM. This current implementation of QSVM is still limited, as it can only classify linearly separable datasets, and the decision boundary has to cross the origin.

\section*{Acknowledgment}
We would like to thank Wolfgang John, Per Persson and Anton Frisk Kockum for providing feedback on the first draft of the paper.

%

\bibliographystyle{IEEEtran}
\bibliography{IEEEabrv,refqsvm}

\begin{thebibliography}{10}
\providecommand{\url}[1]{#1}
\csname url@samestyle\endcsname
\providecommand{\newblock}{\relax}
\providecommand{\bibinfo}[2]{#2}
\providecommand{\BIBentrySTDinterwordspacing}{\spaceskip=0pt\relax}
\providecommand{\BIBentryALTinterwordstretchfactor}{4}
\providecommand{\BIBentryALTinterwordspacing}{\spaceskip=\fontdimen2\font plus
\BIBentryALTinterwordstretchfactor\fontdimen3\font minus
  \fontdimen4\font\relax}
\providecommand{\BIBforeignlanguage}[2]{{%
\expandafter\ifx\csname l@#1\endcsname\relax
\typeout{** WARNING: IEEEtran.bst: No hyphenation pattern has been}%
\typeout{** loaded for the language `#1'. Using the pattern for}%
\typeout{** the default language instead.}%
\else
\language=\csname l@#1\endcsname
\fi
#2}}
\providecommand{\BIBdecl}{\relax}
\BIBdecl

\bibitem{ahmed2018automated}
J.~Ahmed, T.~Josefsson, A.~Johnsson, C.~Flinta, F.~Moradi, R.~Pasquini, and
  R.~Stadler, ``Automated diagnostic of virtualized service performance
  degradation,'' in \emph{NOMS 2018-2018 IEEE/IFIP Network Operations and
  Management Symposium}.\hskip 1em plus 0.5em minus 0.4em\relax IEEE, 2018, pp.
  1--9.

\bibitem{preskill2018quantum}
J.~Preskill, ``Quantum computing in the nisq era and beyond,'' \emph{Quantum},
  vol.~2, p.~79, 2018.

\bibitem{boser1992training}
B.~E. Boser, I.~M. Guyon, and V.~N. Vapnik, ``A training algorithm for optimal
  margin classifiers,'' in \emph{Proceedings of the fifth annual workshop on
  Computational learning theory}.\hskip 1em plus 0.5em minus 0.4em\relax ACM,
  1992, pp. 144--152.

\bibitem{zurek2003decoherence}
W.~H. Zurek, ``Decoherence and the transition from quantum to
  classical--revisited,'' \emph{arXiv preprint quant-ph/0306072}, 2003.

\bibitem{anguita2003quantum}
D.~Anguita, S.~Ridella, F.~Rivieccio, and R.~Zunino, ``Quantum optimization for
  training support vector machines,'' \emph{Neural Networks}, vol.~16, no. 5-6,
  pp. 763--770, 2003.

\bibitem{rebentrost2014quantum}
P.~Rebentrost, M.~Mohseni, and S.~Lloyd, ``Quantum support vector machine for
  big data classification,'' \emph{Physical review letters}, vol. 113, no.~13,
  p. 130503, 2014.

\bibitem{harrow2009quantum}
A.~W. Harrow, A.~Hassidim, and S.~Lloyd, ``Quantum algorithm for linear systems
  of equations,'' \emph{Physical review letters}, vol. 103, no.~15, p. 150502,
  2009.

\bibitem{suykens1999least}
J.~A. Suykens and J.~Vandewalle, ``Least squares support vector machine
  classifiers,'' \emph{Neural processing letters}, vol.~9, no.~3, pp. 293--300,
  1999.

\bibitem{li2015anexperimental}
Z.~Li, X.~Liu, N.~Xu, and J.~Du, ``Experimental realization of a quantum
  support vector machine,'' \emph{Physical review letters}, vol. 114, no.~14,
  p. 140504, 2015.

\bibitem{nielsen2010quantum}
\BIBentryALTinterwordspacing
M.~Nielsen and I.~Chuang, \emph{Quantum Computation and Quantum Information:
  10th Anniversary Edition}.\hskip 1em plus 0.5em minus 0.4em\relax Cambridge
  University Press, 2010. [Online]. Available:
  \url{https://books.google.se/books?id=-s4DEy7o-a0C}
\BIBentrySTDinterwordspacing

\bibitem{riofrio2011continuous}
C.~A. Riofr{\'\i}o, ``Continuous measurement quantum state tomography of atomic
  ensembles,'' \emph{arXiv preprint arXiv:1111.5627}, 2011.

\bibitem{schuld2017implementing}
M.~Schuld, M.~Fingerhuth, and F.~Petruccione, ``Implementing a distance-based
  classifier with a quantum interference circuit,'' \emph{arXiv preprint
  arXiv:1703.10793}, 2017.

\bibitem{cai2013experimental}
X.-D. Cai, C.~Weedbrook, Z.-E. Su, M.-C. Chen, M.~Gu, M.-J. Zhu, L.~Li, N.-L.
  Liu, C.-Y. Lu, and J.-W. Pan, ``Experimental quantum computing to solve
  systems of linear equations,'' \emph{Physical review letters}, vol. 110,
  no.~23, p. 230501, 2013.

\bibitem{OCR_dataset}
``Optical character recognition,'' \url{https://github.com/damiles/basicOCR},
  accessed April 05, 2019.

\bibitem{UCI-Iris}
``Uci machine learning repository-iris dataset,''
  \url{https://archive.ics.uci.edu/ml/datasets/Iris}, accessed May 31, 2019.

\bibitem{PyQuil}
R.~S. Smith, M.~J. Curtis, and W.~J. Zeng, ``A practical quantum instruction
  set architecture,'' 2016.

\bibitem{Qiskit}
G.~Aleksandrowicz \emph{et~al.}, ``Qiskit: An open-source framework for quantum
  computing,'' 2019.

\bibitem{IBMQexperience}
``Ibm q experience,'' \url{https://www.research.ibm.com/ibm-q/}, accessed July
  04, 2019.

\bibitem{Ignisgithub}
``Qiskit-ignis,'' \url{https://github.com/Qiskit/qiskit-ignis}, accessed July
  06, 2019.

\bibitem{dueck2018optimization}
G.~W. Dueck, A.~Pathak, M.~M. Rahman, A.~Shukla, and A.~Banerjee,
  ``Optimization of circuits for ibm's five-qubit quantum computers,'' in
  \emph{2018 21st Euromicro Conference on Digital System Design (DSD)}.\hskip
  1em plus 0.5em minus 0.4em\relax IEEE, 2018, pp. 680--684.

\bibitem{endres2003new}
D.~M. Endres and J.~E. Schindelin, ``A new metric for probability
  distributions,'' \emph{IEEE Transactions on Information theory}, vol.~49,
  no.~7, pp. 1858--1860, 2003.

\bibitem{majtey2005jensen}
A.~Majtey, P.~Lamberti, and D.~Prato, ``Jensen-shannon divergence as a measure
  of distinguishability between mixed quantum states,'' \emph{Physical Review
  A}, vol.~72, no.~5, p. 052310, 2005.

\end{thebibliography}

\end{document}